\documentclass[final]{article}

\immediate\write18{./getsvninfo.sh}

\usepackage{graphicx}
\usepackage{amsmath}
\usepackage{amssymb}
\usepackage{seceqn}
\usepackage{amsmath} 
\usepackage{rotating} 
\usepackage{latexsym} 
\usepackage{amscd}
\usepackage{delarray}
\usepackage[sort]{natbib}
\usepackage{afterpage}
\usepackage{color}
\usepackage{amsfonts}
\usepackage{dsfont}

\newcommand{\acom}[1]{\relax}
\newcommand{\bcom}[1]{\relax}

\newcommand{\Rn}[1]{\ensuremath{\mathds{R}^#1}}

\newcommand{\eps}{\varepsilon}

\newcommand{\rf}[1]{(\ref{#1})}

\newcommand{\hpdo}[3]{\ifthenelse{#3 = 2}{{#1}_{#2#2}}{\frac{\partial^{#3} #1}{{\partial #2}^{#3}}}}
\newcommand{\be}{\begin{equation}}
\newcommand{\ee}{\end{equation}}
\newcommand{\bea}{\begin{eqnarray}}
\newcommand{\eea}{\end{eqnarray}}
\newcommand{\beas}{\begin{eqnarray*}}
\newcommand{\eeas}{\end{eqnarray*}}
\newcommand{\sbea}{\begin{subequations}}
\newcommand{\seea}{\end{subequations}}
\newcommand{\bal}{\begin{align}}
\newcommand{\eal}{\end{align}}
\newcommand{\nn}{\nonumber}
\newcommand{\mywhere}{\quad\mbox{where}\quad}
\newcommand{\myand}{\quad\mbox{and}\quad}
\newcommand{\mywith}{\quad\mbox{with}\quad}

\begin{document}
\title{Stationary solutions of driven fourth- and sixth-order 
Cahn-Hilliard type equations}
\author{M. D. Korzec\thanks{Corresponding author, Weierstrass Institute for Applied Analysis and Stochastics (WIAS), D-10117 Berlin, Germany ({korzec@wias-berlin.de}).}
        \and P. L. Evans\thanks{Institute for Mathematics,
 Humboldt University of Berlin, D-10099 Berlin, Germany ({pevans@mathematik.hu-berlin.de}).} \and A. M\"unch\thanks{School of Mathematical Sciences,
University of Nottingham, UK ({andreas.muench@nottingham.ac.uk}).} \and B. Wagner\thanks{Weierstrass Institute for Applied Analysis and Stochastics (WIAS), D-10117 Berlin, Germany ({wagnerb@wias-berlin.de}).}}
\date{\today}
\maketitle
\begin{abstract}
New types of stationary solutions of a one-dimensional 
driven sixth-order Cahn-Hilliard
type equation that arises as a model for epitaxially growing
nano-structures such as quantum dots, are derived by an extension
of the method of matched asymptotic expansions that retains exponentially
small terms. This method yields analytical expressions for far-field
behavior as well as the widths of the humps of these spatially non-monotone solutions
in the limit of small driving force strength which is the deposition rate in case 
of epitaxial growth. These solutions extend the family of the 
monotone kink and antikink solutions. The hump spacing is 
related to solutions of the Lambert $W$ function.

Using phase space analysis for the corresponding fifth-order dynamical
system, we use a numerical technique that enables the efficient and
accurate tracking of the solution branches, where the asymptotic solutions
are used as initial input.

Additionally, our approach is first demonstrated for the related but simpler driven
fourth-order Cahn-Hilliard equation, also known as the convective
Cahn-Hilliard equation.
\end{abstract}


\section{Introduction}

A paradigm for phase separating systems such as binary alloys is 
the Cahn-Hilliard equation for the phase fraction $u$ 
\be
u_t+\left(Q(u)+\eps^2 u_{xx}\right)_{xx}=0\label{ch}
\ee
where $Q(u)$ is the derivative of the double-well potential ${\cal F}$, 
typically 
\be
Q(u)= {\cal F}^\prime(u) = u-u^3.\label{Q}
\ee
The long-time dynamics are characterized by the logarithmically slow 
coarsening process of phases, corresponding to local minima of the potential, separated by interfaces of width $\eps$. 
This process is well described by the motion of equidistantly spaced 
smoothed shock solutions or {\em kinks} (``positive kinks'') and {\em antikinks} (``negative kinks'') which connect the local minimum of ${\cal F}(u)$ at $u=-1$ to that at $u=1$ and vice versa. 

In recent years, an extension of this model has been studied 
for the case when the phase separating system is driven by an external field \cite{L90, YRHJ92}. 
In one space dimension it can be written as 
\be
u_t- \nu u u_x+\left(Q(u)+\eps^2 u_{xx}\right)_{xx}=0 , \label{cch}
\ee 
where $\nu$ denotes the strength of the external field. This equation, the convective Cahn-Hilliard (CCH) equation, also arises as a model for the evolution of the 
morphology of steps on crystal surfaces \cite{SU96}, and the growth of 
thermodynamically unstable crystal surfaces into a melt with 
kinetic undercooling and strongly anisotropic 
surface tension \cite{LM93, gurtin93, GDN98}. 

The dynamics of this model as $\nu\to 0$ are characterized by 
coarsening, as is typical for the Cahn-Hilliard equation ($\nu=0$)  
\cite{EB96, WORD03}.
If $\nu\to\infty$ using the transformation $u\mapsto u/\nu$ in \rf{cch} 
one obtains  
the Kuramoto-Sivashinski equation, which is a well-known model 
for spatio-temporal chaotic dynamics (see e.g. \cite{GNDZ01} and references therein). 
Recently, Eden and Kalantarov \cite{EK07} also established the existence of a finite-dimensional inertial manifold for the CCH equation, viewed as an infinite-dimensional dynamical system. 

A related higher-order evolution equation arises in the 
context of epitaxially growing thin films (for a review on self-ordered nano-structures on crystal surfaces see Shchukin and Bimberg \cite{SB99}). 
Here, the formation of {\em quantum dots} and their faceting  
has been described by the sixth-order equation   
\be
u_t- \nu u u_x-\left(Q(u)+\eps^2 u_{xx}\right)_{xxxx}=0, \label{hcch}
\ee
where $u$ denotes the surface slope, $\nu$ is 
proportional to deposition rate \cite{SGDNV03} and $Q(u)$ is given with \rf{Q}, it is assumed to have this
form from now on throughout the paper. The high order derivatives are a result of the additional regularization energy 
which is required to form an edge between two plane surfaces with 
different orientations. This implies that the crystal surface tension 
also depends on curvature, which becomes very high at edges as the parameter 
$\eps$ goes to zero. 
In analogy to the Cahn-Hilliard equation, 
here the phases are the orientations of the facets. 
This {\em higher-order convective} Cahn-Hilliard (HCCH) equation  
shares many properties with the CCH equation. 
In both cases the dynamics are described by conserved order 
parameters if $\nu=0$. They also share characteristic coarsening dynamics 
as $\nu \to 0$ and chaotic dynamics as $\nu$ gets large. 
To understand the complicated structure of the solutions 
it is instructive to study first the stationary solutions 
and their stability as it has been done for the CCH equation \cite{ZPNG06, L90}. 
For small $\nu$ the stationary solutions for both equations 
have been characterized by the monotone {\em kink} and {\em antikink} 
solutions \cite{L90, SGDNV03}. Recently new spatially non-monotone solutions 
were found for the lower order equation \cite{ZPNG06}. 
In this study we establish that the HCCH equation also possesses such 
non-monotone solutions. We show this by using phase-space methods for the 
corresponding fifth-order boundary value problem. We use 
the expression ``simple'' or ``monotone'' for a solution that 
connects the maximal value of $u(x)$ to the minimal value without any humps on the way down, 
although these extrema exist and lead to non-monotonicity even for simple 
(anti-)kink solutions of the HCCH equation. 

Since the treatment of this high-order problem is not straightforward, 
one part of this study is concerned with the development of an approach 
that accurately locates the heteroclinic connections 
in the five-dimensional phase space. 
We find that these stationary solutions develop 
oscillations whose width and amplitude increase as $\nu\to 0$. 

In the second part of this study we derive an analytic 
expression for the width and amplitude within the asymptotic regime 
of small external field strength. For the 
CCH equation we find that the width has a logarithmic dependency on the strength of the 
external field, while for the HCCH equation our analysis yields a dependency in terms 
of the Lambert $W$ function.
In order to arrive at these expressions we solve the 
fifth-order equation by a combination of the method of matched 
asymptotic expansions and exponential asymptotics. We first demonstrate our 
approach for the third-order boundary value problem arising from the CCH equation. 
Our approach generalizes the work by Lange \cite{lange83} to higher-order 
singularly perturbed nonlinear boundary-value problems, where standard 
application of matched asymptotics is not able to locate the position of 
interior layers that delimit the oscillations of the non-monotone solutions. 

Reyna and Ward \cite{RW95} previously developed an approach to resolve the 
internal layer structure of the solutions to the 
boundary-value problem for the related Cahn-Hilliard and 
viscous Cahn-Hilliard equations. 
The approach is based on a method due to Ward \cite{ward92} who uses a ``near''
solvability condition for the corresponding linearized problem in his 
asymptotic analysis, and who was inspired by an earlier variational method  
\cite{KKM87} and work by O'Malley \cite{omalley76} and Rosenblat et al.\cite{RS80}, who 
investigated the problem of spurious solutions to singular perturbation 
problems of second-order nonlinear boundary-value problems \cite{CP68}. 
Moreover, for the related Kuramoto-Sivashinsky equation, a multiple-scales analysis 
of the corresponding third-order nonlinear boundary-value problem 
by Adams et al. \cite{AKT03} shows that the derivation of 
monotone and oscillating traveling-wave  
solutions involve exponentially small terms; their method is based on 
an analysis of the {\em Stokes phenomenon} of the corresponding problem 
in the complex plane (see Howls et al. \cite{HKT99} for an introduction). 

In what follows we begin with the phase space analysis for the 
CCH equation in section 2, followed by the asymptotic treatment for 
$\nu \ll 1$. The asymptotic ideas used for the CCH equation are then 
applied to the HCCH equation in section 3. 
The solutions obtained there are useful to serve 
as initial input for the numerical investigations of 
the branches of non-monotone solutions in section 4.
In this part we develop our numerical approach and then use it to identify new stationary 
solutions of the HCCH equation, these agree with the asymptotic theory. Finally we shortly sum the 
results up together with concluding remarks in the last section 5.

\section{Stationary solutions of the convective Cahn-Hilliard equation}

The high-order term in the CCH equation represents the regularization of  
the internal layers of the solutions. For most of our investigations 
we consider the problem in the scaling of the internal layers, 
or {\em inner} scaling, where the $x$-coordinate is stretched about the 
location $x=\bar x$ of a layer according to 
\be
x^*=\frac{x-\bar x}{\varepsilon}.\label{stretch}
\ee
In this scaling the CCH equation becomes (after dropping the ``$*$'')
\be
\varepsilon^2 u_t- \frac{\delta}{2} (u^2)_x+\left(Q(u)+ u_{xx}\right)_{xx}=0 , \label{cch-inner}
\ee 
where $\delta=\varepsilon\nu$. 
The stationary problem obtained by setting $u_t$ to zero can be integrated once, requiring that the 
solutions approach the constants $\pm \sqrt{A}$ as $x\to\mp \infty$, where $A$ is a constant of integration. That is, 
we consider the boundary value problem 
\be
\frac{\delta}{2} \left(u^2-A\right)=\left(Q(u)+ u_{xx}\right)_{x}
\label{CCHstat}
\ee
together with the far-field conditions
\begin{equation} \lim \limits_{x \rightarrow \pm \infty} u = \mp \sqrt{A}  \label{CCHbc}
\end{equation}
and vanishing derivatives in the same spatial limit. We refer to solutions of this system as antikinks. 
Monotone antikinks are known analytically \cite{L90}, while 
recently, non-monotone connections were computed numerically by Zaks et al. \cite{ZPNG06}.  
We now briefly discuss the numerical approach to obtain these 
solutions. Here we concentrate on the regime where $0<\delta\ll 1$ in 
order to compare with the asymptotic solutions derived later on. 
For a bifurcation analysis for larger $\delta$ we refer to \cite{ZPNG06}.



\subsection{Numerical solutions  \label{numcch}}

For the numerical solutions we will work with a rescaled version, where we set $u = \sqrt{A} c$ so that the equilibrium points do not depend on $A$, and for $Q(u)$ given by \rf{Q}, (\ref{CCHstat}) and (\ref{CCHbc}) become
\begin{equation}
(1 - c^2) =  -\frac{2}{\delta\sqrt{A}}(c_{xx} + c - A c^3)_x \, , \quad \lim \limits_{x \rightarrow \pm \infty} c = \mp 1. \label{ODEscaled}
\end{equation}
For this problem we find it most convenient to present a shooting method which enables us to track solution branches in the $(A,\delta)$ parameter plane.
We transform (\ref{ODEscaled}) to a first order system $U^\prime = F(U)$, where $F:\Rn{3} \rightarrow \Rn{3}$
is the function 
\begin{equation}
F_1(U) = U_{2}, \quad F_2(U) = U_{3}, \quad F_3(U) = (3 A (U_1)^2 - 1) U_2 
+ \frac{\delta \sqrt{A}}{2} ( (U_1)^2 - 1) \,.
\label{1stOrderCCH}
\end{equation}
We work in a three dimensional phase space and denote either vectors or whole trajectories therein with capital $U$'s. We use the same notation for two different objects, because it will be clear from the context what is meant. Subscripts indicate the components.

\begin{figure}[h]
\centering  \includegraphics[width=.8\linewidth]{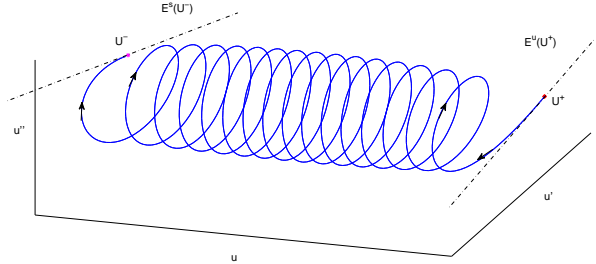} \vspace{-.5cm}
\caption{CCH: Antikink solutions connecting the hyperbolic equilibrium points $U^+$ and $U^-$ are sought in a 3-D phase space. The unstable manifold emerging from $U^+$, $W^u(U^+)$ is one-dimensional, as is the stable manifold $W^s(U^-)$. The approximating linearized spaces $E^u(U^+)$ and $E^s(U^-)$ are drawn as dash-dotted lines and are used in the computations. \label{manifolds}}
\end{figure}

The characteristic polynomials at the equilibrium points $U^\pm = (\pm 1, 0, 0)^T $ are
\begin{equation} {\cal P}^\pm(\lambda) 
= \left|\frac{dF}{dU}(U^\pm) - \lambda I\right| = \lambda^3 + \lambda (1 - 3A) \mp \delta\sqrt{A}  \quad.   \label{eigenvalues}
\end{equation}
The signs of the real parts of the roots determine the dimension of the stable and unstable manifolds $W^u(U^+)$,  $W^s(U^-)$, $W^s(U^+)$,  $W^u(U^-)$ of the equilibrium points. The latter two are two-dimensional and so the existence of a kink is generic, while this is not the case for the antikinks. The dimensions of $W^u(U^+)$ and $W^s(U^-)$ are one, so that the heteroclinic connections from the positive to the negative equilibrium arise from a codimension two intersection. This means that with the two parameters $A$ and $\delta$ we can expect only separated solutions when the manifolds intersect, but due to the reversibility properties which are discussed below the codimension reduces to one and we can expect separated solutions for the free parameter $A$ and a fixed $\delta$, hence one or several whole branches in the $(A,\delta)$ parameter-plane. An example of a non-monotone connection is sketched in figure \ref{manifolds}, where the trajectories wind themselves in the phase space with a solution that exhibits 15 humps.
\paragraph{Reversibility and computations \label{reversible}}
 It is instructive to note that the solution of (\ref{ODEscaled}) is translation invariant, $c(x) \rightarrow c(x + L)$, and forms a reversible dynamical system, hence the solutions are invariant with respect to the transformation $x \rightarrow -x, c \rightarrow -c$, as has also been noted by Zaks et al. \cite{ZPNG06}. 
 
Let us consider generally a $k$-dimensional phase space, since the following discussion will be also useful in section \ref{section:hcchnum} where we analyze the HCCH equation with its higher order system. The linear transformation 
\be
R:\Rn{k} \rightarrow \Rn{k}, \quad R(U_j) = (-1)^j U_j,j=1,\ldots,k \label{eqn:revop}
\ee 
fulfills $R^2 = Id$ and $RF(U) = -F(RU)$ for $k=3$ and \rf{1stOrderCCH} and represents the reversibility in the phase space. It is an involution (or a reflection) and its set of fixed points is the symmetric section of the reversibility, these
are zero at odd components, $U_i=0$ for odd $i$. A solution that crosses such a point necessarily symmetric under $R$, and for each point $U$ on the connection there exists a corresponding transformed point $RU$ somewhere on the branch. In fact there is an equivalence here since odd solutions necessarily cross a point in the symmetric section. It holds that $c$ and its even derivatives have to vanish in the point of symmetry $L$ because of the fulfilled equations $\frac{d^{2m}}{dx^{2m}}c(x+L) = -\frac{d^{2m}}{dx^{2m}}c(-x+L), m=0,1,\ldots, \lfloor k/2 \rfloor,$ and continuity of the solution and its derivatives. 

From the above we conclude that with a shooting method we can stop integrating when we find a point with zero odd components, since the second half of the solution is then given by the set of transformed points under $R$. Hence we define the following distance function for a trajectory $U$ over the interval of integration which helps to find these points
 \begin{equation}
 d_A(U) = \min_x \sqrt{\sum \limits_{i \text{ odd}} U_i(x)^2} \quad .
 \label{distancefctn}
 \end{equation}
 \begin{figure}[h!]
 \centering \hspace{1mm} \includegraphics[width=.7\linewidth]{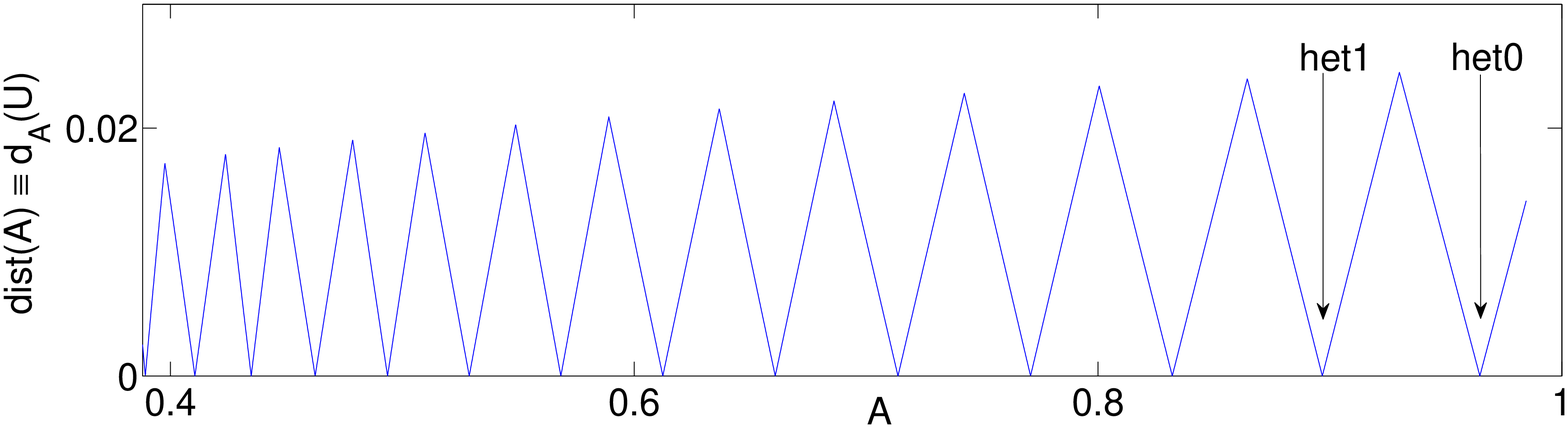}
 \caption{Distance function $d_A$ defined by (\ref{distancefctn}) depending on $A$ with fixed $\delta=0.05$, showing the first 14 zeros corresponding to $het_0$ to $het_{13}$. \label{spikey}}
\end{figure}
\begin{figure}[h!]
 \centering \includegraphics[width=1.05\linewidth]{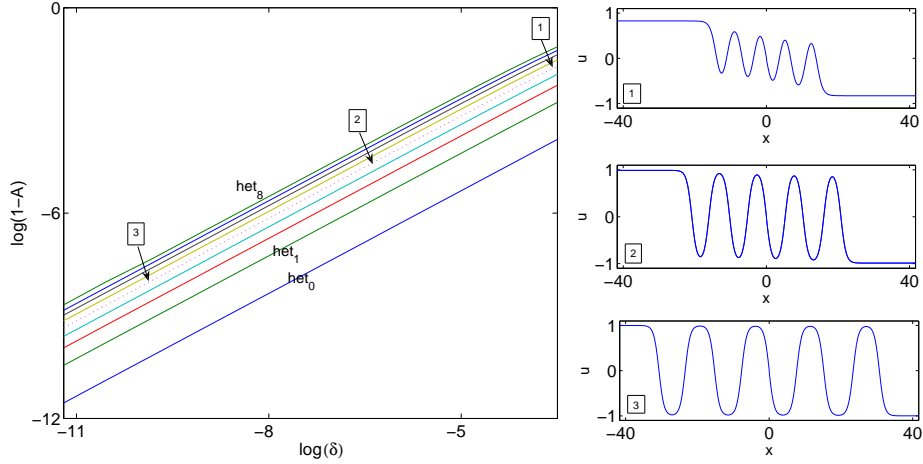} 
 \caption{Parameter plane, $\log(\delta)$ for the x- and $\log(1-A)$ for the y-axis, for the CCH equation for the first 9 antikink solutions $het_k, k=0,1\ldots,8$. The graphs on the right show the shapes of representative $het_4$ solutions, hence those on the fifth line from below, for the approximate $(A, \delta)$ tuples $(0.8259,0.0289)$, $(0.9893,0.0017)$ and $(0.9998, 2.6457 \cdot 10^{-5})$.
\label{parameterplane}}
\end{figure}

The minimization of $d_A(U)$ over the free parameter, $ \min \limits_{A} d_A(U)$, must result in the value zero for an anti-symmetric heteroclinic solution. We can use this condition for shooting and BVP formulations, for both the CCH and later the HCCH equation in section \ref{section:hcchnum}. \\
For a fixed value of $\delta$ and a range of different $A$ we follow the relevant branch of $W^u(U^+)$ by shooting from an initial point $U^+ \pm\epsilon v$ near $U^+$, where $v$ is a unit eigenvector corresponding to the positive eigenvalue of $dF/dU_{\mid U = U^+}$ and $\epsilon \ll 1$. We stop the integration if a certain threshold value for $|U_1|$ is crossed. Figure \ref{spikey} shows $d_A(U)$ as a function of $A$ for $\delta = 0.05$.

At this point we have heteroclinic connections for one fixed value of $\delta$ which we denote by  $het_k, k=0,1,\ldots$ (using the notation in Zaks et al. \cite{ZPNG06}). $het_0$ is the analytical, monotone $tanh$ solution while $het_k$ has $k$ humps on the way down from $\sqrt{A}$ to $-\sqrt{A}$. We will use the same terminology for the solution structure of the stationary HCCH problem in section \ref{section:hcchnum}. 
Here, a $het_k$ solution corresponds to the $k$th zero from the right in figure \ref{spikey}. We then follow the roots of the distance function by linearly extrapolating to a new guess for $A$ and use a bisection algorithm to converge fast to the next root. Figure \ref{parameterplane} shows a portion of the ($A,\delta$) parameter-plane, where we concentrate on very small values of $\delta$, or differently interpreted, on the bifurcation of the various spiraling CCH orbits from the heteroclinic connections of the CH equation in its one dimension smaller phase space.

\subsection{Asymptotic internal layer analysis}

For the asymptotic analysis we use a slightly different scaling than for the 
numerical treatment. Here, we let 
\be
x^*=\frac{x-\bar x}{\sqrt{2}\,\varepsilon}\label{x*}
\ee
denote the {\em inner} variable about a layer located at $x=\bar x$. 
For the stationary problem we then obtain 
\be
\left(u'' + 2\,Q(u)\right)'=\delta\sqrt{2} \left(u^2-A\right)
\label{cchorg}
\ee
instead of \rf{CCHstat}, where $'=d/dx^*$. For later comparisons of the numerical and asymptotic results we have to keep in mind that the spatial scales differ by a factor of $\sqrt{2}$.
 
We point out that
the problem considered here shares the internal layer structure of the 
singular perturbation problems discussed by Lange \cite{lange83}, and 
we will make use and extend this ansatz for our situation. 
This will also prove useful to understand 
the approach taken for the HCCH problem in section \ref{subsec:1humph}, 
since there we have to 
carefully combine the exponential matching with the conventional 
matching procedure when matching the two regions. For both problems the asymptotic analysis can be conveniently carried
out in terms of the small parameter $\delta$.

\begin{figure}[h!]
\centering
\includegraphics*[width=0.7\textwidth]{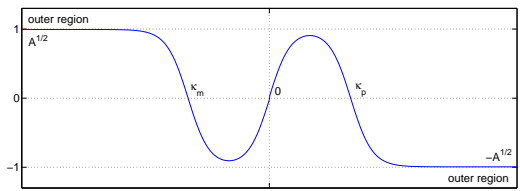}
\caption{\label{fig:1humpsketch}Sketch of a 1-hump, or $het_1$ solution showing the general setup for the matching procedure for the CCH and HCCH equations.} 
\end{figure}

In the following analysis we consider the simplest case of a non-monotone 
solution with only one hump, as illustrated in figure \ref{fig:1humpsketch}; we note that 
non-monotone solutions with more oscillations can be treated similarly. 

\subsubsection{The 1-hump solution}\label{subsec:1hump}

We observe that the 1-hump solution has three internal layers, one at 
$\kappa_m<0$, one at $\kappa_p$ and one at the symmetry point in between. Since the solution is 
point symmetric we can choose this point to be $x=0$ and it will be enough to only discuss the two layers at $\kappa_m$ and zero and then match them to the {\em outer} solution. 
\paragraph{Internal layer near $\boldsymbol\kappa_m$} 

For the first internal layer at $\kappa_m$ we let
\begin{equation}
x_m=\frac{x}{\sqrt{2}\,\varepsilon}- \frac{\bar\kappa_m}{\sqrt{2}}\quad, \label{xm}
\end{equation} 
where $\bar\kappa_m<0$ 
and set
\be
\kappa_m= \bar \kappa_m +\sqrt{2}\sum^{\infty}_{k=1}\delta^k \kappa_{mk}, 
\label{kappa}
\ee
so that to leading order the location where the solution crosses zero is 
$\bar\kappa_m$ and the additional terms account for the corrections 
due to the higher order problems. 

With $u_m(x_m)=u(\eps( \bar\kappa_m + \sqrt{2} \, x_m))$ the governing equation becomes
\be
u_m'''+2\,Q'(u_m)=\delta\,\sqrt{2}\,(u_m^2-A)\,,\mywhere \prime=\frac{d}{dx_m}.\label{cchm}
\ee
For the boundary condition where $u_m$ crosses zero we have  
\be
u_m\left(\frac{\kappa_m-\bar\kappa_m}{\sqrt{2}}\right) = 0
\label{umbc0}
\ee
and the condition towards $-\infty$ is 
\be
\lim_{x_m\to -\infty} u_m(x_m)=\sqrt{A}.
\label{cchmbc0}
\ee

We now assume $u_m(x_m)$ can be written as the following asymptotic expansion, 
valid near $\kappa_m$
\be
u_\alpha(x_\alpha) = u_{\alpha0}(x_\alpha) + \sum_{k=1}^{\infty}\delta^k\, u_{\alpha k}(x_\alpha) \, ,
\label{umexp}
\ee
with $\alpha = m$ here. Additionally, we assume $A$ has the asymptotic expansion 
\be
A= 1 + \delta A_1 + O(\delta^2).\label{Aexp}
\ee
Observe that from \rf{cchmbc0} and \rf{Aexp}
\be
\lim_{x_m\to -\infty} u_m(x_m)= \lim_{x_m\to -\infty} u_{m0}(x_m) + \sum_{k=1}^{\infty}\delta^k\, u_{mk}(x_m) =1+\frac12\sum_{k=1}^{\infty} \delta^k A_k \,. \label{cchmbcinfty}
\ee

To leading order in $\delta$ we get the problem for the Cahn-Hilliard equation 
\sbea\label{leadum}
\begin{align}
&u_{m0}'''+ 2\,Q'(u_{m0})=0\\
&u_{m0}(0)=0\myand \lim_{x_m\to -\infty} u_{m0}(x_m) = 1
\end{align}
\seea
with the unique solution $u_{m0}(x_m)=-\tanh(x_m)$.
Next, the problem of order $\delta$ is 
\sbea\label{um1}
\begin{align}
&\Big({\cal L}\left(u_{m1}, x_m\right)\Big)'=\sqrt{2}\,\left(\tanh^2(x_m)-1\right)
\quad\\
&u_{m1}(0)=\kappa_{m1}\myand \lim_{x_m\to -\infty} u_{m1}(x_m) = \frac{A_1}{2}\,,
\end{align}
\seea
where $\kappa_{m1}$ and $A_1$ are constants to be exponentially matched and the operator ${\cal L}$ is defined by 
\be
{\cal L}(v, z)=v''+ 2\left(1 - 3\,\tanh^2(z)\right)\, v\,,\label{oper}
\ee
and $z=x_m$, $v=u_{m1}$ and $\prime=d/dx_m$. 
Note that the first boundary condition is obtained by expanding \rf{umbc0} 
\bea
u_m\left(\sum_{k=1}^{\infty} \delta^k\kappa_{mk}\right) &=& 
u_m\Big(\,\delta\kappa_{m1}+\delta^2\kappa_{m2} + O(\delta^3)\,\Big)\\
&=& u_{m0}(0)+\delta\big(\,\kappa_{m1} u_{m0}'(0) + u_{m1}(0)\big) +O(\delta^2)\nn
\eea
so that collecting the terms of order $\delta$ gives 
\be
u_{m1}(0)=-\kappa_{m1}\,u_{m0}'(0)=\kappa_{m1}\,.\nn
\ee
Next, we integrate \rf{um1} once to obtain
\be
{\cal L}\left(u_{m1}, x_m\right)= f_m(x_m)\,, \label{eqn:Lf}
\ee
where 
$
f_m(x_m)=-\sqrt{2}\tanh(x_m)+c_m
$. 
Taking the limit of this equation to $-\infty$ yields
$c_m=-\sqrt{2}-2A_1$ so that 
\be
f_m(x_m)=-\sqrt{2}\left(\tanh(x_m)+1\right) - 2A_1\,.
\ee
The homogeneous solutions of (\ref{eqn:Lf}) are 
\bea
\phi_{m}(x_m)&=&-u_{m0}'(x_m)=1-\tanh^2(x_m) \, , \label{phim1}\\
\psi_{m}(x_m)&=&\left(\int^{x_m}_0\frac{dz}{\phi^2_{m}(z)}\right)\phi_{m}(x_m)\,.\label{phim2}
\eea
Also note that $\lim_{x_m\to -\infty}\phi_{m}(x_m)=0$ and $\psi_{m}(0)=0$. 
At this stage it is convenient to choose the inhomogeneous solution that remains bounded as $x_m\to -\infty$ and vanishes at $x_m=0$ which is satisfied by  
\be
\varphi_{\alpha}(x_\alpha)=\psi_{\alpha}(x_\alpha)\int_{-\infty}^{x_\alpha}\,\phi_{\alpha}\,f_\alpha\,dz - \phi_{\alpha}(x_\alpha)\int_{0}^{x_\alpha}\,\psi_{\alpha}\,f_\alpha\,dz\, \label{eqn:inhom}
\ee
with $\alpha = m$. Hence, the unique solution for \rf{um1} 
is the linear combination
\be
u_{m1}(x_m)=-\kappa_{m1}\phi_{m}(x_m)+\varphi_{m}(x_m)\,.\label{solum1}
\ee

\paragraph{Internal layer near $x=0$} 

For the internal layer near the origin we proceed as above. 
Here, we stretch the independent variable as 
\be
x_0=\frac{x}{\sqrt{2}\eps}\label{x0}
\ee
and construct an asymptotic expansion (\ref{umexp})
near $x=0$ with $\alpha = 0$  
for the solution of the problem 
\be
u_0'''+2\,Q'(u_0)=\delta\,\sqrt{2}\,(u_0^2-A)\,,\mywhere \prime=\frac{d}{dx_0}.\label{cch0}
\ee
We note that the point $x=0$ is assumed to be the symmetry point of the complete solution, 
hence here we require 
\be
u_0(0)=0\myand u_{0}''(0)=0\,.\label{u0bc}
\ee
In anticipation of the exponential matching we also require that $\lim_{x_0\to -\infty} u_{00}(x_0) = -1$, so that 
the solution to the leading order problem is 
$u_{00}(x_0)=\tanh(x_0)\label{u00}$
For the solution to $O(\delta)$ we find
\be
u_{01}(x_0)= b_0\,\psi_{0}(x_0)+\varphi_{0}(x_0)\label{u01}
\ee
where $b_0$ is a further constant to be exponentially matched. 
Here, the homogeneous solutions are 
\be
\phi_{0}(x_0) = -u_{00}'(x_0)\myand \psi_{0}(x_0) =  \left(\int^{x_0}_0\frac{dz}{\phi^2_{0}(z)}\right)\phi_{0}(x_0)  \label{phi012}\
\ee
and the inhomogeneous solution is defined by (\ref{eqn:inhom}), where $\alpha = 0$ and $f_0(x_0)=-\sqrt{2}\tanh(x_0)$. They are chosen such that $\varphi_{0}(0)=0$ and $\varphi_{0}''(0)=0$, 
in fact we have $\lim_{x_0\to\pm\infty}\varphi_{0}(x_0)=\pm\sqrt{2}/4$.

\subsubsection{Exponential matching}

Exponential matching requires that all exponentially small and exponentially growing terms have to be accounted for and matched. This means 
first that we have to express the variable $x_0$ in terms of $x_m$ 
(or vice versa). From the definitions of these variables it follows that
\be
x_0=x_m+\frac{\bar\kappa_m}{\sqrt{2}}\,. \label{x0trafo}
\ee
In particular, exponential terms in the solution $u_0(x_0)$ 
transform as $e^{2 x_0} = e^{\sqrt{2}\bar\kappa_{m}}\,e^{2 x_m}$ and so forth for higher order exponential terms $e^{2 n x_0}$ or terms with different signs in the exponent. 

Now note that as $x_0 \rightarrow -\infty$ the leading and $O(\delta)$ solutions can be written as 
\be
u_{00}(x_0)=-1+2 e^{2 x_0} - O(e^{ 4 x_0})
\ee
and with $ \bar \mu = \left(\frac32 b_0+\sqrt{2}\right) $
\be
u_{01}(x_0) = -\frac14 \bar \mu -\frac{b_0}{16}e^{-2 x_0} +\left(\frac{13}{16}b_0+ \frac{1}{\sqrt{2}} + \bar \mu x_0\right)e^{2 x_0}+ O(e^{4 x_0})\,.
\ee
Written in $x_m$ variables the solution 
\bea
u_0(x_m)&=& -1+2e^{2 x_m} e^{{\sqrt{2}\bar\kappa_m}}
+O(e^{2{\sqrt{2}\bar\kappa_m}})\label{237}\\
&&\hspace*{-0.0cm} +\delta\left(-\frac14 \bar \mu -\frac{b_0}{16}e^{-2 x_m} e^{-{\sqrt{2}\bar\kappa_m}}
+(\frac{13}{16}b_0+\frac{1}{\sqrt{2}}+\bar \mu (x_m+\frac{\bar\kappa_m}{\sqrt{2}\eps}))e^{2 x_m} e^{{\sqrt{2}\bar\kappa_m}} \right .\nn\\
&&\left . + O(e^{2{\sqrt{2}\bar\kappa_m}})\right.\bigg)  + O(\delta^2)\nn
\eea
has to be exponentially matched to 
\bea
u_m(x_m) &=& -1+2e^{-2 x_m}+O(e^{- 4 x_m})\label{umexpmatch}\\
&&+\delta\left(-(A_1+\frac{\sqrt{2}}{4}) 
-\frac14 (A_1+\frac{1}{\sqrt{2}} )e^{2 x_m} + (\frac72 A_1+\frac54\sqrt{2}+4\kappa_{m1} )e^{-2 x_m}\right . \nn \\
&&\left . - (3A_1+\frac{1}{\sqrt{2}} )\,x_m e^{-2 x_m}
+O\big(e^{-4 x_m}\big)\right.\bigg) + O(\delta^2)\,\nn
\eea
as $x_m\to\infty$. While we have already anticipated matching of the 
constants during the derivation of the leading order solutions, the constant terms of the $O(\delta)$ solutions are first to be matched. Matching to the exponential terms in \rf{umexpmatch} entails a rearranging of terms of different orders of magnitude in the expansion \rf{237}. In particular, the first exponential term to leading order in \rf{umexpmatch} matches the second term 
of $O(\delta)$ in \rf{237}, the second and largest exponential term  
of $O(\delta)$ in \rf{umexpmatch} matches the second term of the leading order 
in \rf{237}, and so forth. Summarizing, we obtain 
\be
\frac14 (\frac32 b_0 + \sqrt{2} ) = A_1+\frac{\sqrt{2}}{4}, \quad 
 -\rho\frac{b_0}{16} = 2, \quad 
-\frac{\rho}{4} (A_1+\frac{1}{\sqrt{2}} ) = 2 \,, \label{matchm0}
\ee
where we denote 
$
\rho=\delta \, e^{-\sqrt{2}\bar\kappa_m }\,.
$
Solving yields
\be
\rho=4\sqrt{2},\quad A_1=-\frac{3}{\sqrt{2}} \myand b_0=-\frac{8}{\sqrt{2}}\,.
\label{cchA1}
\ee
We observe that we have determined the $O(\delta)$ correction $A_1$. 
Additionally, we now know that $\delta \,e^{-\sqrt{2}\bar\kappa_m}=4\sqrt{2}$, hence
\be
\bar\kappa_m=\frac{\ln\left(\delta\right)}{\sqrt{2}}
-\frac{\ln\left(4\sqrt{2}\right)}{\sqrt{2}}\label{barL1}
\ee
and if we recall \rf{kappa} and $\kappa_m < 0$ 
then the width of the hump is $-\kappa_m$, where
\be
\kappa_m=\frac{\ln\left(\delta\right)}{\sqrt{2}} - \frac{\ln\left(4\sqrt{2}\right)}{\sqrt{2}}
+O\left(\delta\right) \, . \label{L1}
\ee
Further constants, such as $\kappa_{m1}$ are found by including higher exponential terms and expansions of the higher order problems. 
Finally we note that making use of the symmetry of the solution 
about the point $x=0$,  
the exponential matching of the solution near zero to the one 
near $\kappa_p$ proceeds analogously.

\subsubsection{Comparison of numerical and asymptotic solution}

\begin{figure}
 \centering \includegraphics[width=.8\linewidth]{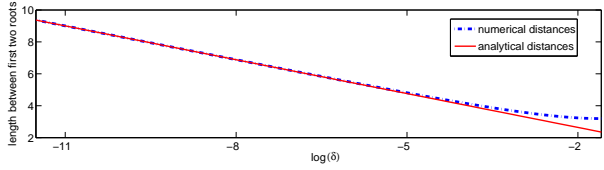} 
 \caption{ Distances between the first two roots of the $het_1$ solutions versus $\log(\delta)$ together with the width predicted by the asymptotic formula \rf{L1}. 
\label{distances_cch} }
\end{figure}

For the comparison with the asymptotic solution we are interested 
mainly in the $het_1$ solution which we derived in section \ref{subsec:1hump}. By numerical continuation of the shooting method, one obtains $N$ tuples $(A^{(j)}, \delta^{(j)}),j=1,\ldots,N$ in the parameter plane that give a $het_1$-branch when being connected. We use two vectors of parameters which we abbreviate ${\bf A} = (A^{(j)})_{j=1,\ldots,N}$ and ${\boldsymbol \delta} = (\delta^{(j)})_{j=1,\ldots,N}$ to confirm the formulas we obtained in the previous section. Further we make use of a distance vector ${\bf K} = (K^{(j)})_{j=1,\ldots,N}$. ${\bf K}$ contains the distances between the zero crossings of the solutions, or in context of the asymptotics section (see figure \ref{fig:1humpsketch}) $K^{(j)} \approx |\kappa_m(\delta^{(j)})|$. To obtain the relation between $A$ and $\delta$ and the evolution of the distances we solve the least squares problems
$$\min \limits_{\mu_1} \| ({\bf 1} - \mu_1 {\boldsymbol \delta}) - {\bf A} \|^2_2 \quad \text{and} \quad \min \limits_{\eta_1, \eta_2} \| \eta_1 \log({\boldsymbol \delta} \eta_2) - {\bf K} \|^2_2 \quad ,$$
hence we assume a linear law for the $A$-values in $\delta$ and a general logarithmic law for the distances.
We obtain 
\be
A \approx 1 - 2.12 \delta \approx 1 - \frac{3}{\sqrt{2}}\delta \myand K \approx -0.71 \log(0.18 \delta) 
\ee 
which confirms the results from the analysis \rf{cchA1} and (\ref{L1}). We see the good match in the distance plot in figure \ref{distances_cch}. 

These results motivated us to obtain a general rule for the relation between the two parameters of the CCH equation for different stationary solutions. The numerically computed branches in figure \ref{parameterplane} 
show that the slopes of the $het_k$ branches are one when plotting $\log(\delta)$ against $\log(1-A)$, so that the relation $\log(\delta) + const = \log(1-A)$ shows the linear dependence $A(k) = 1 + A_1(k) \delta$, where $A(k)$ is the $A$ value for the $het_k$ solution and $A_1(k)$ its linear coefficient. We see that the dependence of $A_1$ with respect to the order of the heteroclinic connection $k$ behaves linearly and we obtain a general expression for the squared far field value of non-monotone $het_k$ solutions, it is given with 
\be 
A_1(k) = -\frac{2k+1}{\sqrt{2}}\,. \label{A1_cch}
\ee

\section{Matched and exponential asymptotics for the stationary HCCH equation}

As for the CCH equation we will perform our analysis of the internal layers in the 
inner scaling \rf{x*}. From the stationary form of \rf{hcch} we obtain 
the equation 
\be
\left(u'' + 2\,Q(u) \right)'''=-\delta\,\, 2^{3/2} \left(u^2-A\right)\,,
\label{hcchorg}
\ee
after integrating once and requiring that for an antikink 
$\lim_{x\to\pm\infty} u=\mp \sqrt{A}$ and setting $\delta = \epsilon^3 \nu$ here.
We consider the $het_1$ (one hump) solution, and 
again make use of the point symmetry of the problem. Now however,
unlike for the CCH equation, the solutions in the {\em outer} region
are not just constants. Here, we have to introduce an {\em
outer} layer to the left of the {\em inner} layer about $\kappa_m$,
see also figure~\ref{fig:1humpsketch} for the case of a 1-hump solution.
In the following subsections we first briefly derive the solution to
this {\em outer} problem and match it to the solution to the {\em inner}
problem near $\kappa_m$. The remaining degrees of freedom are then used
to exponentially match it to a second inner layer near $x=0$.

It has been demonstrated in \cite{SGDNV03} for monotone antikink solutions 
of the HCCH equation, 
that it is necessary to match terms up to order $\delta$ in order to
obtain the correction $A_1$, given 
the asymptotic expansion of $A$ 
\be
A= 1 + \sum^{\infty}_{k=1}\delta^{k/3} \, A_k \,.\label{Aexphcch}
\ee
Here, for the non-monotone antikinks
we have to match {\em inner} and {\em outer} solutions and then also exponentially
match the {\em inner} layers. This has to be carried through iteratively 
up to three orders of magnitude 
in order to obtain not only the correction $A_1$ but also the
expression for the width of the humps.

\subsection{The 1-hump solution for the HCCH equation}\label{subsec:1humph}

We start by shifting to the inner coordinates that describe 
the region 
near $\kappa_m$, which is to be matched to the outer region. 
Again defining $x_m$ by \rf{xm}, the 
governing equation in this inner region is
\be
\Big( u_m''+2\,Q(u_m)\Big)'''= -\,2^{3/2}\,\delta\,(u_m^2-A)\mywhere \prime=\frac{d}{dx_m} \quad .\label{hcchm}
\ee
For the boundary conditions we again place $\kappa_m$ near the point where 
$u_m$ crosses zero, i.e. 
\be
u_m\left(\frac{\kappa_m-\bar\kappa_m}{\sqrt{2}}\right) = 0\,.
\label{humbc0}
\ee
The condition towards $-\infty$ is not as trivial as for the CCH equation but 
needs to be matched to the outer solution in the 
region to the left of $\kappa_m$ (or to the right of $\kappa_p$, 
taking account of symmetry). 

For the {\em outer} region (see figure \ref{fig:1humpsketch}), 
where $x_m$ becomes very large, we use the ansatz 
\be
\xi=\delta^{1/3} \, x_m \myand Y(\xi;\delta)=u_m(x_m;\delta)\,\label{xi}
\ee
and obtain the {\em outer} problem 
\be
\Big(\delta^{2/3}\,Y_{\xi\xi}+2\,Q\left(Y\right)\Big)_{\xi\xi\xi} = -\,2^{3/2}\,\left(Y^2-A\right)\label{equU}
\ee
with the far field condition
\be
\lim_{\xi\to -\infty} Y(\xi)=\sqrt{A}\label{lim-infty} \, .
\ee

The region near $x=0$, for which we use the variable $x_0$ from \rf{x0},
is described by the problem 
\be
\Big( u_0''+2\,Q(u_0)\Big)'''= -\,2^{3/2}\,\delta\,(u_0^2-A)\mywhere \prime=\frac{d}{dx_0}\label{hu0}
\ee
The point $x=0$ is the point of symmetry of the solution. Here we require 
\be
u_0(0)=0,\quad u_{0}''(0)=0\myand u_{0}''''(0)=0 \,, \label{hu0bc}
\ee
plus additional conditions from the exponential matching to the internal layer 
near $\kappa_m$ as $x_0\to -\infty$, as we have shown for the CCH equation. 

Here we assume the solutions to these three problems for $Y$, $u_m$ and $u_0$ can be represented 
by asymptotic expansions 
 \be
u_{\alpha}(x_{\alpha}; \eps) = u_{{\alpha}0}(x_{\alpha}) + \sum_{k=1}^{\infty}\delta^{k/3}\, u_{{\alpha}k}(x_{\alpha}),
\mywhere \alpha=0, m
\label{hualphaexp}
\ee
valid near $\kappa_m$ and $x=0$, respectively, and 
\be
Y(\xi;\delta)=Y_0(\xi)+\sum^{\infty}_{k=1}\delta^{k/3}\,Y_k(\xi)\,,\label{U}
\ee
valid in the outer region, where we let  
\be
\kappa_m = \bar\kappa_m +\sqrt{2}\sum^{\infty}_{k=1}\delta^{k/3} \kappa_{mk}\,.
\label{hkappa}
\ee

To obtain solutions to the outer problem is straightforward 
\cite{SGDNV03}, but in order to be more comprehensible we 
include the results in appendix~\ref{app-outer-HCCH}. 
The solutions to the other regions are discussed now. 

\subsubsection{Leading order}

To leading order in $\delta$ we get the problem 
\sbea\label{leadhum}
\begin{align}
&\left(u_{m0}''+ 2\,Q(u_{m0})\right)'''=0\label{eqhum0}\\
&u_{m0}(0)=0\label{bchum0}
\end{align}
\seea
Matching to the leading order outer solution \rf{U0} $Y_0=1$ we find  
\be
u_{m0}(x_m)=-\tanh(x_m)\,.\label{hum0}
\ee
Its representation towards the internal layer about $x=0$ is given by 
\be
u_{m0}= -1+2 e^{ -2 x_m}-2 e^{ -4 x_m}+O( e^{ -6 x_m})\label{hum0matched}
\ee
as $x_m\to \infty$. The leading order problem for this region is 
\sbea\label{leadhu0}
\begin{align}
&\left(u_{00}''+ 2\,Q(u_{00})\right)'''=0\label{eqhu00}\\
&u_{00}(0)=0,\quad u_{00}''(0)=0\myand u_{00}''''(0)=0\label{bchu00}
\end{align}
\seea
and its solution is
\be
u_{00}(x_0)=\tanh(x_0) \, . \label{hu00}
\ee
As $x_0\to -\infty$ its behavior is given by 
\be
u_{00}= -1+2 e^{ 2 x_0 }-2 e^{ 4 x_0 } + O( e^{ 6 x_0 })\,.\label{hu00exp}
\ee

\subsubsection{O($\boldsymbol\delta^{1/3}$)}
\paragraph{Internal layer near $\bf x = \boldsymbol\kappa_m$}
The expansion of \rf{hcchm} and \rf{humbc0} to order $\delta^{1/3}$ yields 
\sbea
\begin{align}
&{\cal L}(u_{m1},x_m)=f_{m1}(x_m)\label{eqhum1}\\
&u_{m1}(0)=-u_{m0}'(0)\,\kappa_{m1}=
\kappa_{m1}\label{bchum1}
\end{align}
\seea
where ${\cal L}$ is defined by \rf{oper} as for the CCH equation and
\be
f_{m1}(x_m):=c_{1m}x_m^2+c_{2m}x_m+c_{3m}\,.\label{fm1}
\ee
The homogenous solutions are therefore \rf{phim1} and \rf{phim2}. The constants $c_{1m}, c_{2m}, c_{3m}$ are obtained by three successive integrations of the ODE for $u_{m1}$ obtained at this order.
We choose the inhomogeneous solution so that 
it grows only algebraically as $x_m\to -\infty$ and vanishes at 
$x_m=0$. Particular solutions to \rf{bchum1} are of the form
\be
\varphi_{\alpha j}(x_\alpha) = \psi_{\alpha}(x_\alpha)\int_{0}^{x_\alpha}\,\phi_{\alpha}\,f_{\alpha j}\,dz - \phi_{\alpha}(x_\alpha)\int_{0}^{x_\alpha}\,\psi_{\alpha}\,f_{\alpha j}\,dz +  \gamma_{\alpha j} \psi_{\alpha}(x_\alpha)\,,\label{phimi1}
\ee
so that now we obtain $\varphi_{m1}$ for $\alpha = m, j=1$ in \rf{phimi1} and 
\be
\gamma_{m 1} = -\frac{\pi^2}{12} c_{1m} + \ln(2) c_{2m} - c_{3m} \,.
\ee
Hence the solution is 
\be
u_{m1}(x_m)=-\kappa_{m1}\phi_{m}(x_m)+\varphi_{m1}(x_m)\,.\label{solhum1}
\ee
We evaluate $\psi_\alpha, \phi_\alpha$ etc. and subsequent functions with the assistance of \textit{Maple}. As $x_m\to -\infty$ the limiting behavior of $u_{m1}$ is 
\bea
u_{m1}(x_m)&&=
-\frac{1}{8}(c_{1m} + 2c_{3m}) - \frac14 c_{2m} x_m - \frac14 c_{1m} x_m^2 \label{hum1a}\\
&&+\left( \frac{1}{64}(-7c_{1m}-8c_{3m} + 256 \kappa_{m1} + 30 c_{2m} + 4 c_{2m} \pi^2 - 72 c_{1m}\zeta(3)) \right .\nn\\
&&\left . + \frac{1}{16}(-6 c_{2m} + 15 c_{1m} + 24 c_{3m})x_m + \frac{1}{8}(6c_{2m} - 3 c_{1m})x_m^2 +\frac12 c_{1m} x_m^3 \right) e^{ 2 x_m } \nn \\ 
&& +O(\ e^{ 4 x_m })\nn
\eea
where $\zeta$ is the Riemann Zeta function, and $u_{m1}$
must match the outer solution which is given in the appendix by \rf{Uxm} and has only constant terms to this order. Hence we require $c_{2m}=0$ and $c_{1m}=0$.
The matched solution is now
\bea
u_{m1}^{(m)}(x_m)&&=(1-\tanh^2(x_m))\,\kappa_{m1}\label{hum1matched}\\
&&-\frac{c_{3m}}{16}\Big(-2 e^{6x_m}+4+10 e^{2 x_m} - 12 e^{4 x_m} - 24 x_m e^{2 x_m} \Big) \frac{e^{-2 x_m}}{\left(e^{2 x_m}+1\right)^2}\,,\nn
\eea
where we denote by $u_{m1}^{(m)}$ the solution that is obtained by matching to the outer solution $Y$. 
As we will see later, exponential matching to the inner solution $u_0$, i.e. as 
$x_m\to\infty$, where we find 
\bea
u_{m1}^{(m)}(x_m)&=&
\frac18c_{3m} e^{ 2 x_m }+\frac12c_{3m}+\left(-\frac74c_{3m} + 4\kappa_{m1} + \frac32 c_{3m} x_m\right) e^{ -2 x_m }\nn\\
&&+\left(\frac{11}{4}c_{3m} - 8 \kappa_{m1} - 3 c_{3m} x_m \right) e^{ -4 x_m}+O( e^{ -6 x_m })\,,\nn
\eea
requires also $c_{3m}=0$. Hence, denoting by $u_{m1}^{(e)}$ the solution that 
has been exponentially matched to the inner solution $u_0$ near $x=0$, we obtain 
\be
u_{m1}^{(e)}(x_{m})= \left(1-\tanh^2(x_m)\right)\,\kappa_{m1}\,.\label{hum1expmatched}
\ee

\paragraph{Internal layer near $x=0$}
The $O(\delta^{1/3})$ problem is 
\sbea
\begin{align}
&{\cal L}(u_{01},x_0)=f_{01}(x_0)\,,\label{eqhu01}\\
&u_{01}(0)=0,\quad u_{01}''(0)=0\myand u_{01}''''(0)=0\,,\label{bchu01}
\end{align}
\seea
with 
\be
f_{01}(x_0):=c_{10}x_0^2+c_{20}x_0+c_{30} \, .\label{f01}
\ee
Its general solution reads
\be
u_{01}(x_0)=\varphi_{01}(x_0) + g_1\,\psi_{0}(x_0)\,,\label{hu01}
\ee
where the homogeneous solutions are as before and the inhomogeneous solution 
is given by equation \rf{phimi1} with $\alpha = 0, j=1$ and 
\be
\gamma_0= -\frac{\pi^2}{12} c_{10} +\ln(2) \, c_{20} - c_{30} \, ,
\ee
so that $\varphi_{01}(0)=0$ and $\varphi_{01}$ grows algebraically 
as $x_0\to -\infty$. Furthermore, symmetry requires 
$\varphi_{01}''(0)=0$ and $\varphi_{01}''''(0)=0$, which implies 
$c_{10}=0$ and $c_{30}=0$ 
leading to 
\bea
\varphi_{01}(x_0) &=&
\frac{c_{20}}{16(1+ e^{ -2 x_0} )^2}\Big(1-4x_0
+12\,{\rm dilog}( e^{ 2 x_0 } + 1) e^{ -2 x_0} - e^{ -4 x_0 } + 12 x_0^2 e^{ -2 x_0}
\Big .\nn\\
&&\Big . +\pi^2 e^{ -2x_0}+12x_0 e^{ -4 x_0} - 14x_0 e^{ -2 x_0} - \ln(1+ e^{ -2 x_0}) e^{ 2 x_0} + 8 e^{ -4x_0}\ln(1+ e^{-2x_0}) \Big .\nn\\
&&\Big . - 8\ln(1+ e^{-2x_0})) + e^{ -6x_0}\ln(1+ e^{-2x_0})+2 e^{-6x_0}x_0\Big)\,,
\eea
where ${\rm dilog}$ denotes the dilogarithm function.
The remaining free parameters of $u_{01}$ to be matched are 
$c_{20}$ and $g_1$. As will be demonstrated later, exponential matching 
to $u_m$ requires an expression for $u_{01}$ as $x_0\to -\infty$ 
\bea
u_{01}(x_0)&=& 
-\frac{g_1}{16}e^{-2 x_0}-\frac14c_{20}x_0-\frac38g_1 \\
&&+\frac{1}{32}\Big(2 c_{20} \pi^2 + 15 c_{20} + 26 g_1
+( 48 g_1 - 12 c_{20})x_0+ 24 c_{20} x_0^2\Big)e^{2 x_0}\nn\\
&&+\frac{1}{48} \Big( - 36 g_1 - 89 c_{20}- 6 c_{20} \pi^2 + (84 c_{20} - 144 g_1) x_0 - 72 c_{20} x_0^2\Big)e^{4 x_0} + O(e^{ 6 x_0}) \nn
\eea
and then re-expanding $u_0$ in the variable $x_m$. This shows that 
also $c_{20}=0$, $g_1=0$ and $c_{3m}=0$.
Any other choice leads to a system for the parameters having no solution. Hence, only $\kappa_m$ remains as a free constant in the two regions. 
The exponentially matched solution is therefore simply
\be
u_{01}^{(e)}(x_0)=0\,.\label{hu01final}
\ee
\subsubsection{O($\boldsymbol\delta^{2/3}$)}
\paragraph{Internal layer near $\kappa_m$}
The problem of order $\delta^{2/3}$ is 
\sbea
\begin{align}
&{\cal L}(u_{m2},x_m)=f_{m2}(x_m)\,,\label{eqhum2}\\
&u_{m2}(0)=-u_{m0}'(0)\,\kappa_{m2}-\frac12u_{m0}''\kappa_{m1}^2-u_{m1}'(0)\kappa_{m1}=
\kappa_{m2}-u_{m1}'(0)\,\kappa_{m1}\,,\label{bchum2}
\end{align}
\seea
where 
\be
f_{m2}(x_m):=d_{1m}x_m^2+d_{2m}x_m+d_{3m}+6\,u_{m0}\,(u_{m1}^{(e)})^2\,.\label{fm2}
\ee
Note that ${u^{(m)}_{m1}}'(0)=0$.
Again we choose the inhomogeneous solution so that it grows only algebraically as $x_m\to -\infty$ and vanishes at $x_m=0$ to obtain \rf{phimi1} with $\alpha = m, j=2$ and
\be
\gamma_{m2} = -\frac{\pi^2}{12}d_{1m} + \ln(2) \, d_{2m} - d_{3m} - \kappa^2_{m1} \,,
\ee
so that the general solution is represented as 
\be
u_{m2}(x_m)=-\kappa_{m2}\phi_{m}(x_m)+\varphi_{m2}(x_m)\,.\label{solhum2}
\ee
As $x_m \rightarrow -\infty$ we have to compare 
\bea
u_{m2}(x_m)&=& 
-\frac18 (d_{1m} + 2 d_{3m}) - \frac14 d_{2m} x_m - \frac14 d_{1m} x_m^2 \nn\\
&&+e^{2 x_m} \Big(\frac{1}{64}[(-7 - 72 \zeta(3))d_{1m} - 8 d_{3m}
+256(\kappa_{m2} - \kappa_{m1}^2) + ( 30 + 4 \pi^2 ) d_{2m} ] \Big .\nn\\
&&\Big . + \frac{3}{16}(5 d_{1m} - 2 d_{2m} + 8 d_{3m}) x_m  + \frac{3}{8}(2 d_{2m} - d_{1m}) x_m^2 + \frac{1}{2} d_{1m} x_m^3 \Big) 
+O(e^{4 x_m}) \nn
\eea
with the outer solution. Matching the constant and the linear terms in $x_m$ yields
\be
-\frac14d_{3m}= \frac12A_1-\frac18A_1^2+\frac13C_1A_1+\frac{23}{14}C_1^2+D_1 \, ,
\ee
\be
-\frac14d_{2m}=2^{1/6} C_1 \,.
\ee
There is no quadratic term in the outer solution \rf{Uxm}, 
hence $d_{1m}=0$. 
There are further matching conditions but they do not simplify the 
problem structurally at this point and will be enforced later, 
so that $d_{2m}$, $d_{3m}$ and $\kappa_{m2}$ remain to be determined via 
exponential matching. As $x_m\to \infty$, the expansion to 
this order can be written as
\bea
&& u_{m2}^{(m)} = 
\frac12 d_{3m}-\frac14 d_{2m} x_m + \frac18 d_{3m} e^{2 x_m} + \frac{e^{-2 x_m}}{32} \Big(
- 56 d_{3m}- 15 d_{2m} \big . \label{um2m}\\
&& \big . - 2 d_{2m} \pi^2 + 128 (\kappa_{m1}^2 + \kappa_{m2}) + (48 d_{3m}- 12 d_{2m}) x_m - 24 d_{2m} x_m^2 \Big) + O(e^{-4 x_m}) \, . \nn
\eea

\paragraph{Internal layer near $x=0$}
As for the $O(\delta^{1/3})$ problem, at $O(\delta^{2/3})$ we have 
\sbea
\begin{align}
&{\cal L}(u_{02},x_0)=f_{02}(x_0)\,,\label{eqhu02}\\
&u_{02}(0)=0,\quad u_{02}''(0)=0\myand u_{02}''''(0)=0\,,\label{bchu02}
\end{align}
\seea
with  
\be
f_{02}(x_0):=d_{10}x_0^2+d_{20}x_0+d_{30} \, .\label{f02}
\ee
The general solution is 
\be
u_{02}(x_0)=\varphi_{02}(x_0) + g_2\,\psi_{0}(x_0)\,,\label{hu02}
\ee
where the homogeneous component is as before and the inhomogeneous part is obtained by setting $\alpha = 0, j=2$ and $\gamma_{02} = 0$ in \rf{phimi1}, so that $\varphi_{02}(0)=0$ and $\varphi_{02}$ grows algebraically 
as $x_0\to -\infty$. Symmetry requires $\varphi_{02}''(0)=0$, $\varphi_{02}''''(0)=0$, which implies $d_{10}=0$ and $d_{30}=0$. The remaining free parameters to be matched are $d_{20}$ and $g_2$.
In order to exponentially match to $u_m$ to $O(\delta^{2/3})$ and obtain $u_{m2}^{(e)}$, we  
again have to expand $u_{02}(x_0)$ as $x_0\to -\infty$, giving 
\bea
u_{02}(x_0)&=&
-\frac{\hat \mu}{16}  e^{-2 x_0} - \frac14 d_{20} x_0-\frac38 \hat \mu \\
&&+\frac{1}{32} \Big( (15 + 2  \pi^2 + 2 \ln(2) ) d_{20}  + 26 g_2 +(48 \hat \mu - 12 d_{20}) x_0 
+24 d_{20} x_0^2 \Big) e^{2 x_0} \Big .\nn\\
&&\Big . +\frac{1}{48}\Big( - (89 + 6 \pi^2) d_{20} - 36 \hat \mu 
+ (84 d_{20} - 144 \hat \mu ) x_0 - 72 d_{20} x_0^2 \Big) e^{4 x_0} + O(e^{6 x_0}) \, , \nn
\eea
and re-express in terms of $x_m$, where we have used the abbreviation $\hat \mu = d_{20} \ln(2)+g_2$. 

\subsubsection{O($\boldsymbol\delta$)}
\paragraph{Internal layer near $\kappa_m$}
The problem to be solved at order $O(\delta)$ is 
\sbea
\begin{align}
{\cal L}(u_{m3}, x_m)&=f_{m3}(x_m)\,,\label{eqhum3}\\
u_{m3}(0)&=
-u_{m2}'(0)\kappa_{m1}-u_{m0}''(0)\kappa_{m1}\kappa_{m2}-u_{m0}'(0)\kappa_{m3}\nn\\
&\hspace*{0.5cm}
-\frac16 u_{m0}'''(0)\kappa_{m1}^3-u_{m1}'(0)\kappa_{m2}-\frac12 u_{m1}''(0)\kappa_{m2}^2\,,\label{bchum3}
\end{align}
\seea
with 
\bea
&&f_{m3}(x_m):=2\left((u_{m1}^{(e)})^3+6\,u_{m0}\,u_{m1}^{(e)}\,u_{m2}^{(e)}\right)\label{fm3}\\
&&-2^{3/2}\left[\frac12 {\rm dilog}(e^{2 x_m}+1) + \frac12(1+k_{1m})x_m^2+(\ln(2)+k_{2m})x_m+k_{3m}\right]\,.\nn
\eea
Again we choose the inhomogeneous solution so that it grows only algebraically as $x_m\to -\infty$ and vanishes at $x_m=0$ and so that we obtain $\varphi_{m3}(x_m)$ by using formula \rf{phimi1} with $\alpha = m, j=3$ and $\gamma_{m3} = 0$. 
The solution is 
\be
u_{m3}(x_m)=-u_{m3}(0)\phi_{m}(x_m)+\varphi_{m3}(x_m)\,,\label{solhum3}
\ee
where $k_{1m}$, $k_{2m}$, $k_{3m}$ and $\kappa_{m3}$ remain to be determined via matching.
In order to exclude exponential growth as $x_m\to -\infty$ we obtain the 
relation 
\bea
k_{2m}&=&\frac{\sqrt{2}}{48 \ln(2)}\left(
\kappa_{m1} \left( - ( 12 + 9  \pi^2) d_{2m} + 12 d_{3m} - 24 \kappa_{m2} \right) \right . \nn \\
&& \left . +\sqrt{2} (24 k_{3m}- 12 \ln(2)^2 + k_{1m} \pi^2) \right)\,,
\eea
so that the expansion obtained as $x_m\to -\infty$ is
\bea
u_{m3}(x_m)&=&
\frac{1}{4\sqrt{2}}(1+ k_{1m}+ 4 k_{3m}) + \frac{1}{\sqrt{2}}(\ln(2) + k_{2m}) x_m \label{um3solexp}\\
&& + (k_{1m}+ 1)\frac{\sqrt{2}}{4} x_m^2 + O(e^{2 x_m}) \,.\nn
\eea
Comparing this with the outer solution to $O(\delta)$, equation \rf{Uxm},  yields the matching conditions 
\bea
&&\hspace*{-1cm}\frac{1}{4\sqrt{2}}(1+ k_{1m}+ 4 k_{3m}) = \left(-\frac14 A_1+\frac13 C_1\right) A_2+\left(\frac{7}{12} C_1^2+\frac13 D_1\right) A_1 \label{test1}\\
&&+\frac12 A_3-\frac{59}{216} C_1 A_1^2-\frac{1}{12} 2^{1/3} C_1 + K_1-\frac{23}{7} C_1 D_1+\frac{1}{16} A_1^3+\frac{127}{28} C_1^3\nn
\eea
for the constant terms,  
\be
\frac{1}{\sqrt{2}}(\ln(2) + k_{2m}) = (D_1 - \frac{23}{7} C_1^2) 2^{1/6} \quad \text{and} \quad (k_{1m}+ 1)\frac{\sqrt{2}}{4}= 2^{-2/3} C_1 
\ee
for the linear and the quadratic terms, respectively.\\
Expanding the solution as $x_m\to \infty$ we find 
\bea
u_{m3}(x_m) &=& \frac{1}{192} \Big( \kappa_{m1} d_{2m} (9 \pi^2+ 24)- 48 \kappa_{m1} d_{3m}
+2 \sqrt{2}  \pi^2 (1  - k_{1m}) - 48 \sqrt{2} k_{3m} \Big) e^{2 x_m}\nn\\
&& +\frac{1}{96}\left( \kappa_{m1} d_{2m} (27 \pi^2 + 72) + \sqrt{2} (k_{1m}  (12 - 6 \pi^2) - 96 k_{3m}
-12 +2  \pi^2) \right)
\nn\\
&& + \frac{1}{\sqrt{2}}(\ln(2) + k_{2m}) x_m
+ (k_{1m}+ 1)\frac{\sqrt{2}}{4} x_m^2 + O(e^{-2 x_m})\,,
\eea
and we will exponentially match it to the solution near $x=0$, which we construct next. 

\paragraph{Internal layer near $x=0$}
The general solution to the $O(\delta)$ problem 
\sbea
\begin{align}
&{\cal L}(u_{03},x_0)=f_{03}(x_0)\,,\label{eqhu03}\\
&u_{03}(0)=0,\quad u_{03}''(0)=0\myand u_{03}''''(0)=0\,,\label{bchu03}
\end{align}
\seea
with 
\be
f_{03}(x_0):=-2^{1/2}\left[{\rm dilog}(e^{2 x_0}+1) - {\rm dilog}(2) + 2 \mu_2 x_0 + (1+k_{10})x_0^2\right]\label{f03}
\ee
and the abbreviation $\mu_2 = \ln(2) + k_{20}$ reads
\be
u_{03}(x_0)=\varphi_{03}(x_0) + g_3\,\psi_{0}(x_0)\,,\label{hu03}
\ee
where 
we have required that $u_{03}(0)=0$ and $u_{03}''(0)=0$. If we also 
enforce $u_{03}''''(0)=0$ then $k_{10}=0$. Again we take an 
inhomogeneous solution 
$\varphi_{03}(x_0)$ which satisfies the above conditions, so that the 
general solution is obtained with 
$$
\mu_1 =  \sqrt{2}( \ln(2)^2 + 2 k_{20} \ln(2)) - g_3  \quad \text{and} \quad
\omega=\int^1_0 \frac1z\ln\left(\frac{z^2+1}{2z}\right)^2-\frac{\ln(2z)^2}{z}dz\approx 0.3094
\,,
$$
\bea
u_{03}&=&
\frac{12 \mu_1 - \pi^2 \sqrt{2} }{192}  e^{-2 x_0} 
+\frac{1}{96}( 36 \mu_1 + \sqrt{2}(12-\pi^2))
 + \frac{\mu_2}{\sqrt{2}} x_0+\frac{\sqrt{2}}{4} x_0^2  \\
&&+\bigg[ \frac{1}{192} \left( 156 \mu_1 + \sqrt{2} [ (19 - 24 k_{20})\pi^2 - 15 - 288 \omega - 180 \mu_2] \right) \nn\\
&&+\frac{1}{16} \left( -24 \mu_1 + \sqrt{2}(12 \mu_2 - 11) \right) x_0 + \frac{\sqrt{2}}{8} \left( 3 - 12 \mu_2 \right) x_0^2-\frac{1}{\sqrt{2}} x_0^3
\bigg] e^{2 x_0} +O(e^{4 x_0})\,. \nn
\eea
For exponentially matching to $u_m$ this again has to be re-expressed 
in $x_m$ and combined with the corresponding expressions 
for $u_{00}$, $u_{01}$ and $u_{02}$ . 
This will be done in the next section.
\subsection{Exponential matching}
Now we have to match the rest of the solution $u_m(x_m)$ to the 
rest of the solution $u_0(x_0)$. This requires matching the exponential 
terms in addition to the algebraic terms, similarly to the procedure for the CCH equation, i.e.  
matching of the solution describing the internal layer near $x=\kappa_m$ 
to the solution 
near $x=0$ requires expressing the variable $x_0$ in terms of $x_m$ 
(or vice versa). Recall again that $x_0=x_m+ \bar\kappa_m/\sqrt{2}$ and that
$\bar\kappa_m<0$; the $e^{2 x_0}$ terms in 
the $u_0$ expansion will produce $e^{2 x_m}$ terms with a 
factor $e^{\sqrt{2}\bar \kappa_m}$ (and analogously for $e^{-2 x_0}$ terms) 
and so we will find their corresponding matching partner at a different 
order in $\delta$ in the $u_m$ expansion, as we have shown for the 
CCH equation. The somewhat subtle difference here is that additionally 
we need to determine 
the relationship between $e^{\sqrt{2}\bar\kappa_{m}}$ and $\delta$ 
and we have in principle several choices, only one of which allows a consistent 
matching of both expansions. 
One can observe that the choice 
$
e^{\sqrt{2}\bar\kappa_{m}} = \rho\,\delta^{1/3}\,,
$
where $\rho$ is some constant quickly leads to a contradiction. 
However, setting 
\be
e^{\sqrt{2}\bar\kappa_{m}} = \rho\,\delta^{2/3}\label{shift}
\ee
will lead to a $O(\delta^{2/3})$ shift of terms, so that e.g. 
\be
e^{2 x_0}\quad\mbox{will shift to a term}\quad \delta^{2/3}\, e^{2 x_m}\,,
\ee
\be
e^{-2 x_0}\quad\mbox{will end up as a term}\quad \delta^{-2/3}\, e^{-2 x_m}
\ee
and so forth, so that e.g. a term $e^{2 x_0}$ in the leading order 
part of the $u_0$ expansion will have to match a $e^{2 x_m}$ term in 
the $O(\delta^{2/3})$ part of the $u_m$ expansion, or a $e^{-2 x_0}$ term 
in the $O(\delta)$ part of the $u_0$ expansion will have to match a 
$e^{-2 x_m}$ term in the $O(\delta^{1/3})$ part of the $u_m$ expansion. 
This will also produce terms that will have no partner term in the 
transformed expansion. Their coefficients must then be set to zero. 
If we now sum the expansions for $u_{01}(x_0)$, $u_{02}(x_0)$ 
and $u_{03}(x_0)$ and re-expand using \rf{shift}, we obtain 
\bea
u_0(x_m)&=& 
-1-\frac{1}{16}\Big( d_{20} \ln(2)+g_2\Big) e^{-2 x_m} \rho
+\frac{1}{192} \bigg( 12 \mu_1 - \sqrt{2} \pi^2 \bigg) e^{-2 x_m}\, \rho \,\delta^{1/3}\nn\\
&&+\frac{1}{24}\bigg( d_{20} (3 \ln(\rho) - 9 \ln(2)- 2 \ln(\delta) )- 9 g_2 - 6 d_{20} x_m + 48 e^{2 x_m}/\rho \bigg) \delta^{2/3}  \nn\\
&&+\bigg[\frac{1}{96} \left(36 \mu_1 + \sqrt{2} [ 12 + (16 \ln(\delta) - 24 \ln(\rho)) \mu_2 + 6(\ln(\rho) - \frac{2}{3}\ln(\delta))^2 - \pi^2 ] \right) \nn\\
&&+ \frac{\sqrt{2}}{12} \left(2 \ln(\delta) + 6 \mu_2 - 3\ln(\rho) \right) x_m +\frac{\sqrt{2}}{4}  x_m^2\bigg] \delta \,,
\eea
which has to match $u_{m1}(x_m)$, $u_{m2}(x_m)$ 
and $u_{m3}(x_m)$ to each order, respectively. From this we obtain 
further conditions for the parameters in addition to those we have 
already found. Solving the complete system of equations 
then yields the solutions for the width of the hump 
\be
\Delta = \frac{\sqrt{2}}{6} \ln\left(\frac{\beta}{W(\beta^{1/3})^3}\right)\,,  \label{Delta}
\ee
with $\beta = 2^{11}/(27\delta^2)$, where $W$ is the Lambert $W$ function 
(so $W(x)$ is the solution of $x = W \exp(W)$). The expressions for the remaining matching constants $C_1, D_1$, etc. are omitted. The first correction in \rf{Aexphcch} has the coefficient 
\be
A_1= - 3\, 2^{1/6}\,. \label{A1-hcch}
\ee
Note that in the transformed expansions as well as in the expressions for 
the parameters also contain so-called {\em logarithmic switch-back} terms.

\section{Numerical method for the fifth-order phase space \label{section:hcchnum}}

\begin{figure}[h!] 
\centering  \includegraphics[width=.65\linewidth]{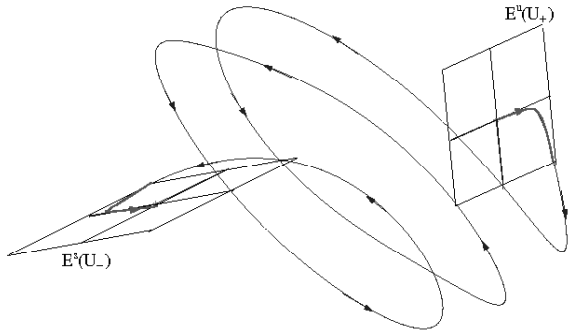} 
\caption{HCCH: Heteroclinic orbits between the equilibrium points are sought in a 5-D phase space that is indicated here in 3D. The manifolds $W^u(U^+)$ and $W^s(U^-)$ are two-dimensional which is suggested by the two planes in the picture. \label{HCCHmanifolds}}
\end{figure}

For the numerical stationary solutions of the HCCH equation (\ref{hcchorg}) we apply the same scaling for $u$ that we used for the CCH equation to obtain equilibrium points at $\pm 1$.
\be
(1 - c^2) =  \frac{2}{\delta\sqrt{A}}(c_{xx} + c - A c^3)_{xxx} \,, 
\quad \lim \limits_{x \rightarrow \pm \infty} c = \mp 1 \quad, \label{HCCHscaled}
\ee
again assuming that derivatives vanish in the far field. Reduction to a first order system $U^\prime = F(U)$, with $F:\Rn{5} \rightarrow \Rn{5}$, gives a five-dimensional phase space, where the first
four components of $F_i(U)$ are equal to $U_{i+1}$ and the fifth is 
\be 
F_5(U)
= 6 A (U_2)^3 + 18 A U_1 U_2 U_3 + (3 A (U_1)^2 - 1) U_4 + \delta \sqrt{A} ( 1 -
(U_1)^2)/2 \,.
\ee 
The equilibrium points are $U^\pm =
\pm (1, 0, 0, 0, 0)^T $ and at these points the characteristic polynomials are
\begin{equation} 
{\cal P}^\pm(\lambda) = \lambda^5 + \lambda^3 (1 - 3A) \pm \delta \sqrt{A} \,.  \label{HCCHeigenvalues}
\end{equation}
For small $\delta$ the manifolds $W^u(U^+)$ and $W^s(U^-)$ are both two-dimensional, resulting in a codimension two event when searching for heteroclinic solutions connecting the two hyperbolic fixed points $U^+$ and $U^-$. The HCCH equation exhibits the same reversibility properties as its lower order version. This reversibility is again given by the transformation \rf{eqn:revop} from the CCH section, which also here fulfills $RF(U) = -F(RU)$. The codimension reduces by one and again we deal with a codimension one problem and two parameters, hence we may expect solution branches in the $(A,\delta)$ parameter plane. Section \ref{numcch} showed that a condition for the existence of heteroclinic orbits is a value where the distance function (\ref{distancefctn}) reaches zero and the same condition holds for the HCCH equation.
The phase space is sketched in figure \ref{HCCHmanifolds}, indicating the linearizations of the intersecting manifolds in the equilibrium points.
For this problem a shooting method will be very slow and may lead to bad accuracy since the 
additional parameter, say $\varphi \in [0 , 2 \pi)$, an 
angle defining points on a circle close to the equilibrium point on the linearization of the two-dimensional manifold, requires a very fine resolution to obtain heteroclinic solutions.

 
\subsection{Boundary value problem formulation \label{bvp_sec}}
There exist several possibilities to set up equations for finding heteroclinic connections in a boundary value problem framework. Generally one crucial stumbling block is the choice of a suitable phase condition that picks a certain solution out of the infinitely many available ones due to phase shifts \cite{FD91, B90}. We choose to incorporate one phase condition proposed by Beyn \cite{B90}, for which we use
an approximation of the solution, $V$, typically given by a previous solution for slightly different parameter values. Equation \rf{HCCHscaled} contains two parameters, $A$, $\delta$, and in
addition the truncated domain length $L$. As discussed by Doedel et al. \cite{DF89} one of
the free parameters can be replaced by $L$ to find a
connection. We replace $\delta$, solve and continue after extrapolating to an
approximate value of $A$ for a nearby chosen and fixed $\delta$. Rescaling
the domain to $[0, 1]$ yields, with the phase condition variable $U_{ph}$ introduced by Beyn \cite{B90} the first order system
\sbea\begin{align}
  U_i^\prime  & = L U_{i+1}, \quad i=1,2,3,4 \\ 
  U_5^\prime & = L \left( 6 A (U_2)^3 + 18 A U_1 U_2 U_3 + (3 A (U_1)^2 - 1) U_4 + \delta \sqrt{A} \frac{( 1 - (U_1)^2)}{2} \right) \\
  U_{ph}^\prime & = L (V^\prime)^T U   \\ 
  L^\prime & = 0, \,  A^\prime  = 0 \quad .
\end{align}\seea
Hence, we obtain one equation for the phase condition and two for the parameters in addition to the five given by the original ODE, i.e., we have an overall system of eight equations which have to be supplemented by the same number of boundary conditions. At the edges of the domain we utilize {\em projected boundary conditions} \cite{HW80, B90},  which make use of eigenvectors in the equilibrium points and can be incorporated by computing $V_0$, the matrix whose columns are composed by the eigenvectors which correspond to the eigenvalues at the upper equilibrium point $U^+$ with negative real part, and by forming the counterpart $V_1$ containing those eigenvectors given by the unstable directions at the lower stationary point $U^-$. Hence, we consider the eight boundary conditions
\begin{align}
  U_{ph}(0) = 0, \quad U_{ph}(1) = 0, \quad  V_0^T(U(0) - U^+) = 0, \quad  V_1^T(U(1) - U^-) = 0 \,.
\end{align}
For initial estimates we can use solutions obtained from the asymptotic analysis of section \ref{subsec:1humph}, i.e. the leading order solution $\tanh$ profiles $$V(x) = -\tanh(x-K) + \tanh(x) - \tanh(x+K) \quad ,$$ 
for the $het_1$ solution with guessed root-distance $K$.

\begin{figure}[h!]
\centering  \includegraphics[width=.9\linewidth]{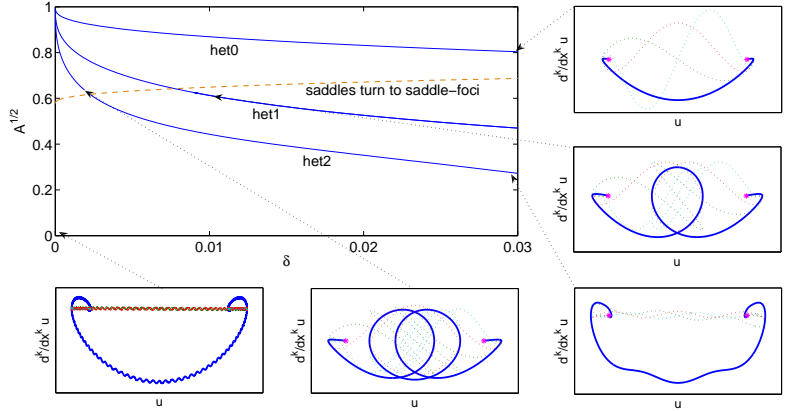} 
\caption{$(\sqrt{A},\delta)-$plane with curves for the first three heteroclinic connection branches for the HCCH equation. The dashed line in the parameter plane indicates the position where the positive roots of the characteristic polynomial in $U^+$ have nonzero imaginary parts. Below and to the right we see five phase space diagrams (tuples $(U_1, U_2),(U_1, U_3), \ldots$) for selected solutions pointed out with arrows marking the corresponding parameters. The first pair $(U_1, U_2)$ is plotted as bold solid curve.   \label{AD_hcch}}
\end{figure}

The boundary value solvers we use are based on mono-implicit Runge-Kutta formulae \cite{SKF05, KS01}. As for the CCH problem efficiency can be improved by making use of the theory from section \ref{reversible} which holds analogously for the HCCH equation to obtain a boundary condition at the fixed point of a point-symmetric solution. We can use half of the previous domain length and phase conditions become redundant, because the phase is already fixed. We replace the projected boundary conditions by 
$$U_1(0) = 1, \quad U_2(0)^2 + U_3(0)^2  = 0, \quad U_4(0)^2 + U_5(0)^2  = 0$$
so that together with the self-reversibility condition on the right interval end $U_1(1) = U_3(1) = U_5(1)= 0$ we have six conditions which match the five equations together with the free parameter $A$.
Final solutions are obtained by reflecting the solution and its derivatives around zero and changing the signs of the first, third and fifth component. 
Examples of branches of different solutions are shown in 
figure \ref{AD_hcch}. 
\subsection{Solutions and comparison to analytical results}
\begin{figure}[h!]
\centering  \includegraphics[width=.6\linewidth]{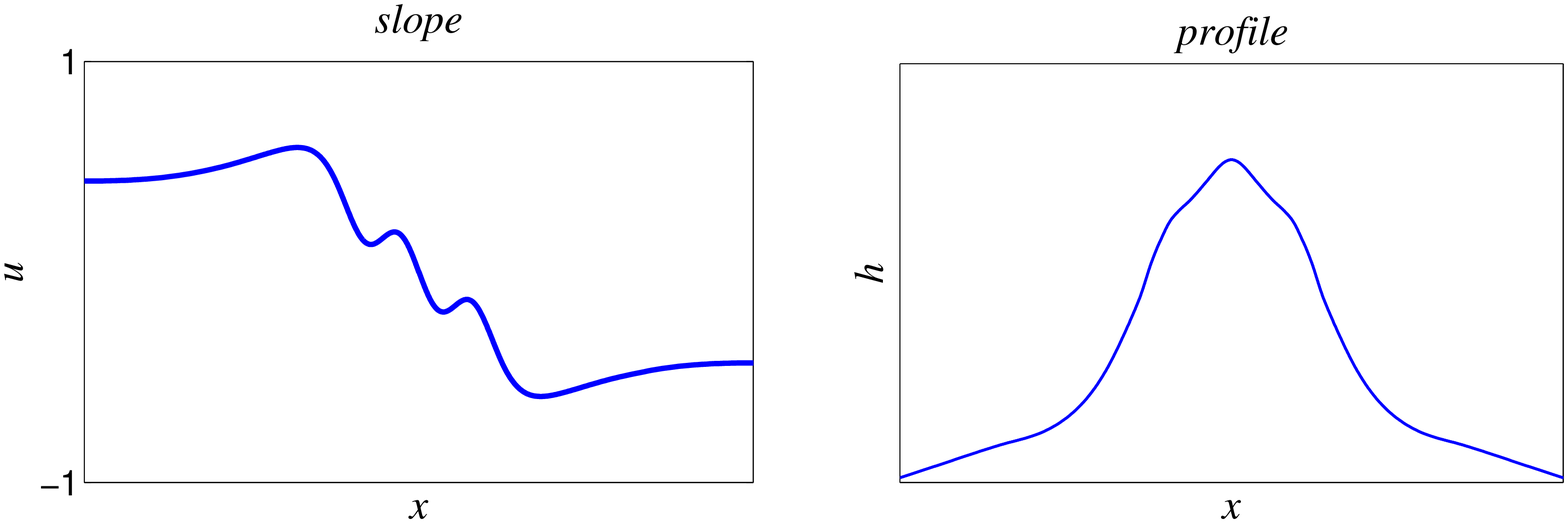} 
\caption{$het_2$ solution for $\delta = 0.01$ and $A = 0.443$ and the corresponding profile obtained by integration. \label{hcch-het2}}
\end{figure}

With the boundary value formulation we are able to compute new HCCH stationary solutions. In figure \ref{hcch-het2} we see a particular $het_2$ solution and the profile of the growing structure. 

Up to 3D one can nicely visualize heteroclinic orbits in the corresponding phase space, while when the dimension is four or higher and the derivatives vanish in the far field one can still plot the 2D phase spaces $(U_1, U_2), (U_1,U_3),\ldots$ and demand connections between the equilibrium tuples $(\pm \sqrt{A},0)$ as a necessary condition for heteroclinic orbits in the higher order space. Several such projections onto 2D are shown in figure \ref{AD_hcch}, where we also see a very rapidly oscillating heteroclinic curve in the bottom left plot which was found by a shooting approach with a minimization procedure that used the two parameters and an angle as free parameters and the distance function (\ref{distancefctn}) as objective function, depending on those parameters. It indicates that as shown for the CCH equation we can in fact find many more $het_k$ branches than those presented for $k=0,1,2$, all emerging from $(A, \delta) = (1,0)$, which corresponds to the Cahn-Hilliard equation.

\begin{figure}[h!]
\centering  \includegraphics[width=1\linewidth]{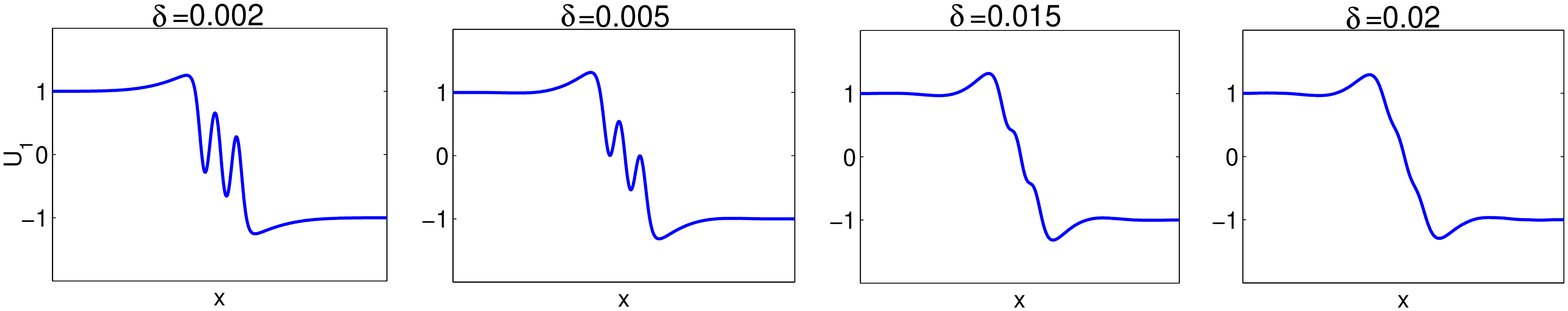} 
\caption{Structural change of the scaled $het_2$ solution as $\delta$ is increased. \label{het2_hcch}}
\end{figure}

In figure \ref{het2_hcch} we see the change in appearance of solutions on the $het_2$ branch as $\delta$ is increased. The shape varies from a solution with two pronounced humps to a monotone one, similar to the $het_0$ solution, although associated with different, smaller, values of $A$. This is crucial if one wants to compute solutions for bigger $\delta$ with a boundary value solver. It easily happens that the solver switches between solution branches, however, this can be prevented by starting continuation in a parameter regime where the high-slope parts of the solutions are non-monotone, and continuing with small steps.
A characteristic of the HCCH solutions is the overshoot from the equilibrium value before the solutions go down. This is not observed for the CCH equation, where the shape is similar at these regions to hyperbolic tangent functions. 

\begin{figure}[h!]
\centering  \includegraphics[width=1\linewidth]{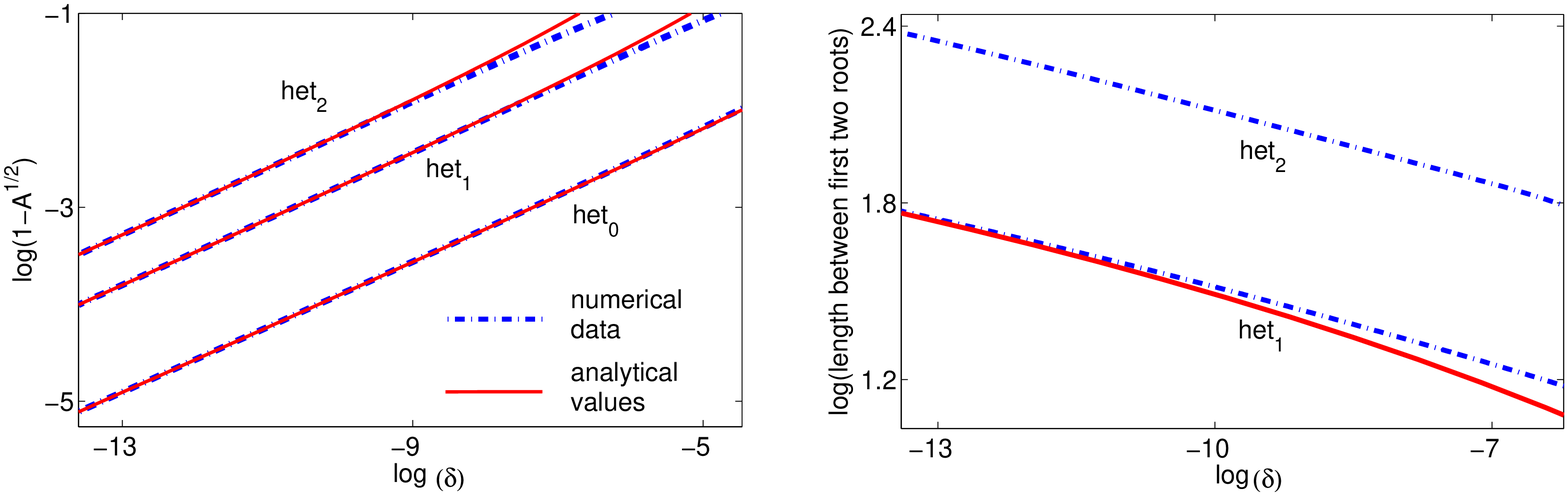}\,\,\, 
\caption{Left figure: Logarithmic version of the $(\sqrt{A}, \delta)$ plot for very small $\delta$. Drawn through curves giving the analytical values, dash-dotted lines those computed with the BVP solver. On the right we see the distances between the first two roots of the $het_1, het_2$ solutions, numerically and for $het_1$ via the analytical expression \rf{Delta} (solid line). \label{AD_log}}
\end{figure}

In light of the expansion \rf{Aexphcch} we try to estimate the ${\cal O}(\delta^{1/3})$ terms $A_1$ for the different heteroclinic connections in a range of very small $\delta$. As we see in figure \ref{AD_log} on the left, the numerically obtained values for $A$ behave like $A = 1 - 2^{1/6} \delta^{1/3}$ in case of the $het_0$ solutions, so that $A_1 = -2^{1/6}$, which is consistent with the result in Savina et al. \cite{SGDNV03}. The numerical result for $het_1$ is in line with the analytical value \rf{A1-hcch} and since for $het_2$ we see the agreement $A_1 \approx -5\,2^{1/6}$, we propose for higher order trajectories that for $het_k$ we have the general approximation $A_1 \approx -(2k+1)\,2^{1/6}$, which is reminiscent of the CCH expression \rf{A1_cch}. Hence this formula is used in figure \ref{AD_log} to plot the \textit{analytical values}.  

We measure the distance between the first and second root for the $het_1$ and the $het_2$ solutions as seen in figure \ref{AD_log} on the right. We compare this to the analytical expression \rf{Delta} for the one-hump solutions in the same figure and see that for small $\delta$ the agreement is good. For both $het_1$ and $het_2$ solutions the distance is seen to increase logarithmically as $\delta$ decreases.

\section{Conclusion}
We have demonstrated that a sixth-order generalization
of the convective Cahn-Hilliard
equation admits multiple stationary solutions connecting constant values.
As for the fourth-order convective Cahn-Hilliard solution,
these include a simple base solution, which is monotone
for the CCH and ``almost'' monotone for the HCCH equation. More complex
solutions, containing multiple humps, are also possible
for each value of the forcing parameter $\delta$,
given particular values of the integration constant $A(\delta)$.
These non-monotone stationary solutions constitute an essential
part of the solution structure for this higher-order Cahn-Hilliard
type equation.
We have demonstrated this via a numerical investigation
of the phase space in which we are able to follow solution
branches. For the simplest of the multi-humped solutions,
the $het_1$ branch,
careful use of matched asymptotics that accounts
for exponentially small terms allows us to find a
solution which yields both the length scale for
the solution (the ``hump length'')
and the parameter value $A(\delta)$
at which it occurs, in the limit of small $\delta$.
Extension of the analysis to higher branches appears
feasible.
Our numerical evidence suggests that similarly
simple asymptotic expressions hold for these
branches, for both the CCH and HCCH equations.
%
%
Physically, these solutions may represent situations where
the edge energy regularization represented in \rf{hcch}
fails to produce a smooth transition between facets.
%
%

Various issues, such as the stability of these solutions
 are presently being considered in the light of applications 
of the HCCH equation as a model for the morphology and dynamics of
quantum dots. In particular, how do adjacent internal layers
 derived from these solutions interact, and
what is their effect on 
the coarsening behavior in large spatial domains?
Savina et al. \cite{SGDNV03} have begun an investigation
of these questions by numerical simulation of \rf{hcch};
it is likely that asymptotics can yield further insights.
 
Physically, further interesting questions relate to the extension
of the HCCH model to richer models for the energetics
of facetted surfaces, and analyzing the three-dimensional extension of the model.

\section*{Acknowledgments}
This work was performed as part of Project C-10 of the DFG research
center {\sc Matheon}, Berlin. AM also gratefully acknowledges the support 
from the Heisenberg Fellowship of the DFG (grant MU 1626/3). 
\begin{appendix}
\section{Outer Problem}\label{app-outer-HCCH}

For the solution to the outer problem \rf{equU}, \rf{lim-infty} 
it is easy to observe that to leading order in $\delta$ the solution of 
\be
Q\left(Y_0\right)_{\xi\xi\xi} =-\sqrt{2}\left(Y_0^2-1\right)\mywith 
\lim_{\xi\to -\infty} Y_0(\xi)=1\label{equU0}
\ee
is 
\be
Y_0(\xi)=1\,.\label{U0}
\ee
To $O(\delta^{1/3})$ the general solution to the problem 
\be
Y_{1_{\xi\xi\xi}}-\sqrt{2}\,Y_1=-\frac{A_1}{\sqrt{2}}\mywith 
\lim_{\xi\to -\infty} Y_1(\xi)=\frac{A_1}{2}\label{equU1}
\ee
is
\be
Y_1(\xi)=\frac{A_1}{2}+C_1e^{2^{1/6}\xi}+ e^{-\xi/2^{5/6}}\left[
C_2\cos\left(\sqrt{3}\,\xi/2^{5/6}\right) + 
C_3\sin\left(\sqrt{3}\,\xi/2^{5/6}\right)\right] \, ,
\label{U1general}
\ee
with $C_1, C_2$ and $C_3$ being constants of integration. The far field condition 
requires that $Y_1$ remains bounded as $\xi\to -\infty$. Hence, $C_2=C_3=0$ and 
\be
Y_1(\xi)=\frac{A_1}{2}+C_1 e^{2^{1/6}\xi}\label{U1}
\ee
Using this and the far field conditions, the solution to the 
$O(\delta^{2/3})$ problem 
\be
Y_{2_{\xi\xi\xi}}-\sqrt{2}\,Y_2=-\frac{A_2}{\sqrt{2}}-\frac12\left(3\left(Y_1^2\right)_{\xi\xi\xi}-\sqrt{2}\,Y_1^2\right)\mywith 
\lim_{\xi\to -\infty} Y_2(\xi)=\frac{A_2}{2}-\frac{A_1^2}{8}\label{equU2}
\ee
is
\be
Y_2(\xi)=\frac{A_2}{2}-\frac{A_1^2}{8}+D_1 e^{2^{1/6}\xi}
+\frac{A_1 C_1}{3} e^{2^{1/6}\xi} \left(1-2^{1/6}\xi\right)
-\frac{23}{14}C_1^2 e^{2^{7/6}\xi}\label{U2}
\ee
and to the $O(\delta)$ problem 
\bea
Y_{3_{\xi\xi\xi}}-\sqrt{2}\,Y_3 &=& -\frac{A_2}{\sqrt{2}}+\frac{Y_{1_{\xi\xi\xi\xi}}}{4}+\sqrt{2}\,Y_1 Y_2 
-\frac12\left(Y_1^3+6\, Y_1 Y_2\right)_{\xi\xi\xi} \nn\\
&&\mywith 
\lim_{\xi\to -\infty} Y_3(\xi)=\frac{A_3}{2}-\frac{A_1 A_2}{4}+\frac{A_1^3}{16}
\label{eqU3}
\eea
it is
\bea
Y_3(\xi)&=&\frac{A_3}{2}-\frac{A_1 A_2}{4}+\frac{A_1^3}{16}\label{U3}\\
&&+\bigg[K_1-\frac{2^{1/3}}{12}C_1+\frac13\left(A_1D_1+A_2C_1\right) -\frac{59}{216}C_1A_1^2\bigg.\nn\\
&&\bigg.+\left(\frac{\sqrt{2}}{12}C_1-\frac{2^{1/6}}{3}\left(A_1D_1+A_2C_1\right)+\frac{17}{72}2^{1/6}A_1^2C_1\right)\,\xi +\frac{2^{1/3}}{18}A_1^2C_1\,\xi^2\bigg] e^{2^{1/6}\xi }\nn\\
&&+\bigg[-\frac{23}{7}C_1D_1+\left(\frac{7}{12}+\frac{23}{21}2^{1/6}\xi\right)A_1C_1^2\bigg] e^{2^{7/6}\xi}
+\frac{127}{28}C_1^3 e^{2^{1/6}3\xi} \, , \nn
\eea
with another integration constant $K_1$. Finally, we obtain the asymptotic representation in terms of $x_m$: 

\bea
Y(x_m)&=& 1
+\left[C_1+\frac12 A_1\right]\,\delta^{1/3}
+\left[C_1\,2^{1/6}\,x_m-\frac18\,A_1^2+\frac13\,C_1\,A_1+D_1-\frac{23}{14}C_1^2+\frac12\,A_2\right]\,\delta^{2/3}\nn\\
&&+\left[-\frac{23}{7}C_1^2\,2^{1/6}x_m+D_1\,2^{1/6}x_m+\frac12\,C_1\,2^{1/3}x_m^2+\left(-\frac14\,A_1+\frac13\,C_1\right)A_2  \right .\nn\\
&&\left .+\left(\frac{7}{12}C_1^2+\frac13\,D_1\right)A_1+\frac12\,A_3-\frac{59}{216}C_1\,A_1^2-\frac{1}{12} \,2^{1/3}\,C_1\right .\nn\\
&&\left .+K_1-\frac{23}{7}C_1\,D_1+\frac{1}{16}A_1^3+\frac{127}{28}C_1^3\right]\,\delta \quad .
\label{Uxm}
\eea

\end{appendix}
\bibliographystyle{abbrv}
\bibliography{paper}

\begin{thebibliography}{10}

\bibitem{AKT03}
K.~L. Adams, J.~R. King, and R.~H. Tew.
\newblock Beyond-all-orders effects in multiple-scales asymptotics:
  travelling-wave solutions to the {Kuramoto}-{Sivashinsky} equation.
\newblock {\em J. Engr. Math.}, 45:197--226, 2003.

\bibitem{B90}
W.-J. Beyn.
\newblock The numerical computation of connecting orbits in dynamical systems.
\newblock {\em IMA Journal of Numerical Analysis}, 9:379--405, 1990.

\bibitem{CP68}
G.~Carrier and C.~Pearson.
\newblock {\em Ordinary differential equations}.
\newblock Blaisdell, Waltham, Massachusetts, 1968.

\bibitem{HW80}
F.~de~Hoog and R.~Weiss.
\newblock An approximation theory for boundary value problems on infinite
  intervals.
\newblock {\em Computing}, 24:227--239, 1980.

\bibitem{DF89}
E.~Doedel and M.~Friedman.
\newblock Numerical computation of heteroclinic orbits.
\newblock {\em Journal of Computational and Applied Mathematics}, 26:155--170,
  1989.

\bibitem{EK07}
A.~Eden and V.~K. Kalantarov.
\newblock The convective {Cahn}-{Hilliard} equation.
\newblock {\em Applied Mathematics Letters}, 20(4):455--461, 2007.

\bibitem{EB96}
C.~L. Emmott and A.~J. Bray.
\newblock Coarsening dynamics of a one-dimensional driven {Cahn}-{Hilliard}
  system.
\newblock {\em Phys. Rev. E}, 54:4568--4575, 1996.

\bibitem{FD91}
M.~Friedman and E.~Doedel.
\newblock Numerical computation and continuation of invariant manifolds
  connecting fixed points.
\newblock {\em SIAM J. Num. Analysis}, 28(3):789--808, 1991.

\bibitem{GDN98}
A.~A. Golovin, S.~H. Davis, and A.~A. Nepomnyashchy.
\newblock A convective {Cahn}-{Hilliard} model for the formation of facets and
  corners in crystal growth.
\newblock {\em Phys. D}, 122(1-4):202--230, 1998.

\bibitem{GNDZ01}
A.~A. Golovin, A.~A. Nepomnyashchy, S.~H. Davis, and M.~A. Zaks.
\newblock Convective {Cahn}-{Hilliard} models: From coarsening to roughening.
\newblock {\em Phys. Rev. Lett.}, 86:1550--1553, 2001.

\bibitem{gurtin93}
M.~E. Gurtin.
\newblock {\em Thermomechanics of Evolving Phase Boundaries in the Plane}.
\newblock Clarendon Press, Oxford, UK, 1993.

\bibitem{HKT99}
C.~J. Howls, T.~Kawai, and Y.~Takei, editors.
\newblock {\em Toward the exact WKB analysis of differential equations, linear
  or nonlinear}.
\newblock Kyoto University Press, Kyoto, 1999.

\bibitem{KKM87}
W.~Kath, C.~Knessl, and B.~Matkowsky.
\newblock A variational approach to nonlinear singularly perturbed boundary
  value problems.
\newblock {\em Stud. Appl. Math.}, 77:61--88, 1987.

\bibitem{KS01}
J.~Kierzenka and L.~Shampine.
\newblock A {BVP} solver based on residual control and the {MATLAB} {PSE}.
\newblock {\em ACM Transactions on Mathematical Software}, 27(3):299--316,
  Sept. 2001.

\bibitem{lange83}
C.~G. Lange.
\newblock On spurious solutions of singular perturbation problems.
\newblock {\em Stud. Appl. Math.}, 68:227--257, 1983.

\bibitem{L90}
K.-T. Leung.
\newblock Theory on morphological instability in driven systems.
\newblock {\em Journal of Statistical Physics}, 61:345--364, 1990.

\bibitem{LM93}
F.~Liu and H.~Metiu.
\newblock Dynamics of phase separation of crystal surfaces.
\newblock {\em Phys. Rev. B}, 48:5808--5817, 1993.

\bibitem{omalley76}
R.~E. O'Malley, Jr.
\newblock Phase-plane solutions to some singular perturbation problems.
\newblock {\em J. Math. Anal. Appl.}, 54(2):449--466, 1976.

\bibitem{RW95}
L.~G. Reyna and M.~J. Ward.
\newblock Metastable internal layer dynamics for the viscous {Cahn}-{Hilliard}
  equation.
\newblock {\em Methods Appl. Anal.}, 2:285--306, 1995.

\bibitem{RS80}
S.~Rosenblat and R.~Szeto.
\newblock Multiple solutions of nonlinear boundary-value problems.
\newblock {\em Stud. Appl. Math.}, 63:99--117, 1980.

\bibitem{SU96}
Y.~Saito and M.~Uwaha.
\newblock Anisotropy effect on step morphology described by
  {Kuramoto}-{Sivashinsky} equation.
\newblock {\em J. Phys. Soc. Jpn.}, 65:3576--3581, 1996.

\bibitem{SGDNV03}
T.~V. Savina, A.~A. Golovin, S.~H. Davis, A.~A. Nepomnyashchy, and P.~W.
  Voorhees.
\newblock Faceting of a growing crystal surface by surface diffusion.
\newblock {\em Phys. Rev. E}, 67:021606, 2003.

\bibitem{SKF05}
L.~F. Shampine, P.~H. Muir, and H.~Xu.
\newblock A user-friendly {Fortran} {BVP}-solver.
\newblock {\em Journal of Numerical Analysis, Industrial and Applied
  Mathematics}, 1(2):201--217, 2006.

\bibitem{SB99}
V.~A. Shchukin and D.~Bimberg.
\newblock Spontaneous ordering of nanostructures on crystal surfaces.
\newblock {\em Rev. Modern Phys.}, 71(4):1125--1171, July 1999.

\bibitem{ward92}
M.~J. Ward.
\newblock Eliminating indeterminacy in singularly perturbed boundary value
  problems with transition invariant potentials.
\newblock {\em Stud. Appl. Math.}, 87:95--134, 1992.

\bibitem{WORD03}
S.~Watson, F.~Otto, B.~Rubinstein, and S.~Davis.
\newblock Coarsening dynamics of the convective {Cahn}-{Hilliard} equation.
\newblock {\em Phys. D}, 178:127--148, 2003.

\bibitem{YRHJ92}
C.~Yeung, T.~Rogers, A.~Hernandes-Machado, and D.~Jasnow.
\newblock Phase separation dynamics in driven diffusive systems.
\newblock {\em J. Statist. Phys.}, 66:1071--1088, 1992.

\bibitem{ZPNG06}
M.~A. Zaks, A.~Podolny, A.~A. Nepomnyashchy, and A.~A. Golovin.
\newblock Periodic stationary patterns governed by a convective
  {Cahn}-{Hilliard} equation.
\newblock {\em SIAM Journal on Applied Mathematics}, 66(2):700--720, 2006.

\end{thebibliography}
\clearpage
\addtocounter{tocdepth}{2}
\clearpage
\end{document}